\documentclass[11pt]{article} 
\usepackage{fullpage}
\usepackage{graphicx}
\usepackage{upref}
\usepackage{enumerate}
\usepackage{latexsym}
\usepackage{pdfpages}
\usepackage[bookmarks]{hyperref}
\usepackage{color,graphics}
\usepackage{comment} 
\usepackage{caption}
\usepackage{subcaption}
\usepackage{wrapfig}
\usepackage{braket}

\usepackage{amssymb}
\usepackage{amsmath}
\usepackage{amsthm}
\usepackage{mathrsfs}
\usepackage{mathtools}

\usepackage{epsfig,graphicx}
\usepackage{algorithmic}
\usepackage[ruled,linesnumbered]{algorithm2e}
\usepackage{amsfonts}
\usepackage{tikz} 
\usepackage{varioref}
\usepackage[ansinew]{inputenc}
\usepackage{qcircuit}
\usepackage{authblk}
\usepackage{multirow}
\usepackage{lineno}

\theoremstyle{plain}
\newtheorem{theorem}{Theorem}[section]
\theoremstyle{plain}
\newtheorem{lemma}{Lemma}[section]
\theoremstyle{plain}

\theoremstyle{plain}

\theoremstyle{plain}
\newtheorem{conjecture}{Conjecture}
\theoremstyle{plain}

\theoremstyle{definition}

\newtheorem{definition}{Definition}[section]
\theoremstyle{definition}
\newtheorem{fact}{Fact}[section]

\theoremstyle{remark}

\theoremstyle{definition}

\AtBeginDocument{}

\DeclareMathOperator{\nat}{\mathbb{N}}

\newcommand{\intg}{\mathbb{Z}}

\newcommand{\poly}{\text{poly}}

\newcommand{\tr}{\text{Tr}}

\newcommand{\id}{\mathbb{I}}

\newcommand{\cliff}{\mathcal{C}}
\newcommand{\pauli}{\mathcal{P}}
\newcommand{\clifft}{\mathcal{J}}

\newcommand{\X}{\text{X}}
\newcommand{\Y}{\text{Y}}
\newcommand{\Z}{\text{Z}}
\newcommand{\had}{\text{H}}
\newcommand{\T}{\text{T}}
\newcommand{\CNOT}{\text{CNOT}}
\newcommand{\phase}{\text{S}}

\newcommand{\chan}[1]{\widehat{#1}}
\newcommand{\tcount}{\mathcal{T}}

\newcommand{\sde}{\text{sde}}
\newcommand{\ham}{\text{ham}}

\newcommand{\upath}{\text{Path}}

\newcommand{\tset}{\overline{T}}
\newcommand{\genv}{\mathbb{V}}

\begin{document}
\title{A (quasi-)polynomial time heuristic algorithm for synthesizing T-depth optimal circuits}

\author[1,2]{Vlad Gheorghiu \thanks{vlad.gheorghiu@uwaterloo.ca}}
\author[1,2,3,4]{Michele Mosca \thanks{michele.mosca@uwaterloo.ca}}
\author[1,3]{Priyanka Mukhopadhyay \thanks{mukhopadhyay.priyanka@gmail.com, p3mukhop@uwaterloo.ca (Corresponding author)}}

\affil[1]{Institute for Quantum Computing, University of Waterloo, Waterloo ON, Canada}
\affil[2]{softwareQ Inc., Kitchener ON, Canada}
\affil[3]{Dept. of Combinatorics and Optimization, University of Waterloo, Waterloo ON, Canada}
\affil[4]{Perimeter Institute for Theoretical Physics, Waterloo ON, Canada}

\date{}

\maketitle

\begin{abstract}
 We investigate the problem of synthesizing T-depth optimal quantum circuits over the Clifford+T gate set. 
 First we construct a special subset of T-depth 1 unitaries, such that it is possible to express the T-depth-optimal decomposition of any unitary as product of unitaries from this subset and a Clifford (up to global phase). The cardinality of this subset is at most $n\cdot 2^{5.6n}$.
  We use nested meet-in-the-middle (MITM) technique to develop algorithms for synthesizing provably \emph{depth-optimal} and \emph{T-depth-optimal} circuits for exactly implementable unitaries. Specifically, for synthesizing T-depth-optimal circuits, we get an algorithm with space and time complexity $O\left(\left(4^{n^2}\right)^{\lceil d/c\rceil}\right)$ and $O\left(\left(4^{n^2}\right)^{(c-1)\lceil d/c\rceil}\right)$ respectively, where $d$ is the minimum T-depth and $c\geq 2$ is a constant. This is much better than the complexity of the algorithm by Amy~et~al.(2013), the previous best with a complexity $O\left(\left(3^n\cdot 2^{kn^2}\right)^{\lceil \frac{d}{2}\rceil}\cdot 2^{kn^2}\right)$, where $k>2.5$ is a constant. 
We design an even more efficient algorithm for synthesizing T-depth-optimal circuits. The claimed efficiency and optimality depends on some conjectures, which have been inspired from the work of Mosca and Mukhopadhyay (2020). To the best of our knowledge, the conjectures are not related to the previous work. Our algorithm has space and time complexity $\poly(n,2^{5.6n},d)$ (or $\poly(n^{\log n},2^{5.6n},d)$ under some weaker assumptions). 

\end{abstract}



\section{Introduction}
\label{sec:intro}

The notion of a quantum computer was introduced by Feynman \cite{1982_F} as a solution to the limitations of conventional or classical computers. In numerous fields algorithms designed for quantum computers outperform their classical counterparts. Some examples include integer factorization \cite{1999_S,1994_S}, searching an unstructured solution space \cite{1996_G}. One of the most widely used methods for describing and implementing quantum algorithms is quantum circuits, which consists of a series of elementary operations dictated by the implementing technologies. 

Circuit synthesis and optimization is a significant part of any computer compilation process whose primary goal is to translate from a human readable input (programming language) into instructions that can be executed directly on a hardware. In quantum circuit synthesis the aim is to decompose an arbitrary unitary operation into a sequence of gates from a universal set, which usually consists of Clifford group gates and at least one more non-Clifford gate \cite{2004_AG}. The non-Clifford gates are more expensive to implement fault-tolerantly than Clifford gates. A popular universal fault-tolerant gate set is the ``Clifford+T'', in which the cost of fault tolerant implementation of the T gate \cite{2005_BK, 2009_FSG,2006_AGP} exceeds the cost of the Clifford group gates by as much as a factor of hundred or more in most error correction schemes.
Fault tolerant designs and quantum error correction are essential in order to deal with errors due to noise on quantum information, faulty quantum gates, faulty quantum state preparation and faulty measurements. In particular, for long computations, where the number of operations in the computation vastly exceeds the number of operations one could hope to execute before errors make negligible the likelihood of obtaining a useful answer, fault-tolerant quantum error correction is the only known way to reliably implement the computation. 

With recent advances in quantum information processing technologies \cite{2012_BSK_,2011_BWC_,2012_CGC_,2012_RGP_} and fault-tolerant thresholds \cite{2012_BAOKM,2009_FSG,2012_FWH}, as scalable quantum computation is becoming more and more viable we need efficient automated design tools targeting fault-tolerant quantum computers.
And minimization of the number of T gates in quantum circuits remain an important and widely studied goal. It has been argued \cite{2012_F,2013_AMMR,2014_AMM, 2016_AdMVMPS, 2020_dMGM} that it is also important to reduce the maximum number of T gates in any circuit path. While the former metric is referred to as the \textbf{T-count}, the latter is called the \textbf{T-depth} of the circuit. 

An $n$-qubit quantum circuit consisting of Clifford+T gates implements a $2^n\times 2^n$ unitary. In the context of reducing resources (such as T gates) necessary to implement a unitary $U$, two types of problems have been investigated - (a) \emph{synthesis} and (b) \emph{re-synthesis}. The input to an algorithm for a \textbf{quantum circuit synthesis} problem is a $2^n\times 2^n$ unitary matrix and the goal is to output a circuit implementing it \cite{2006_DN, 2013_GS}. When we impose additional constraints like \emph{minimizing}
certain resources such as T-count or T-depth \cite{2013_AMMR}, we often call this as \textbf{(resource)-optimal synthesis} problem. From here on, we focus on the T-depth as the resource being minimized.  
To be more precise, there can be more than one (equivalent) circuits implementing $U$. A T-depth-optimal synthesis algorithm is required to output a circuit with the \emph{minimum} T-depth. We call this a \textbf{T-depth-optimal circuit}. 
With a slight abuse of terminology, we use the terms ``synthesis algorithm'' and ``T-depth optimal synthesis algorithm'' interchangeably, which should be clear from the context.
It must be observed that with the addition of this tighter constraint on the output (i.e. that it be T-depth optimal), there is a probability that the complexity of the problems change. For example, it was known that a quantum circuit can be synthesized in $\poly(2^n)$ time, where $2^n$ is the input size \cite{2006_DN, 2020_dBBVA}. The work in \cite{2020_MM} was the first to propose a $\poly(2^n)$ time algorithm for synthesizing T-count-optimal circuits.

With an input size $O(2^n)$, we cannot hope to get an optimal synthesis algorithm with complexity less than that. This makes  these algorithms practically intractable after a certain value of $n$. Hence \textbf{re-synthesis} algorithms have been developed, where some more information is provided as input, usually a circuit implementing $U$ \cite{2014_AMM, 2020_HS} and the task is to \emph{reduce} (not minimize) the T-depth in the input circuit. In the literature, nearly every re-synthesis algorithm (usually with complexity $\poly(n)$) does not account for the complexity of generating the initial input circuit from U. This step itself has complexity $O(2^n)$. A full study comparing these two kinds of algorithms and quality of their results is beyond the scope of this work.

Despite their higher complexity compared to re-synthesis algorithms, the importance of studying optimal synthesis algorithms cannot be undermined. They can be used to assess the quality of a re-synthesis algorithm, for example, how close are their output to an optimal one. They can be used to generate the input circuit of a re-synthesis algorithm. A large circuit can be fragmented and the unitary of each part can be synthesized optimally, giving an overall reduction in resources. From a theoretical viewpoint, they shed light on the complexity of problems that are usually harder than their relaxed re-synthesis counterpart. As an illustration for the significance of developing resource-optimal synthesis algorithms we observe the following. In our paper we have been able to generate T-depth-optimal circuits for standard unitaries like Toffoli, Fredkin, Peres and Quantum OR, which were not generated by the re-synthesis methods used in \cite{2013_AMMR}\footnote{It is worth noting that the non-Toffoli unitaries listed here are Clifford equivalent to the Toffoli gate, and thus a T-depth 3 circuit could also be derived from the Toffoli circuit of T-depth 3 discovered in \cite{2013_AMMR}; however the prescribed algorithm failed to find the optimal T-depth for most of the Clifford equivalent unitaries.}. Though this has a T-depth-optimal synthesis algorithm, they could not synthesize beyond 2 -qubit unitaries with T-depth 2. For larger unitaries like the mentioned 3-qubit ones, they used peep-hole optimization, a popular re-synthesis method. Except for the Toffoli, they obtained T-depth 4. The new approach in this paper had significantly lower complexity than the synthesis method in \cite{2013_AMMR} and was able to synthesize T-depth 3 circuits.

The Solovay-Kitaev algorithm \cite{1997_K, 2006_DN} guarantees that given a unitary $U$, we can generate a circuit with a universal gate set like Clifford+T, such that the unitary $U'$ implemented by the circuit is at most a certain distance from $U$ (the distance being induced by some appropriate norm). 
In fact it has been proved that we can get a Clifford+T circuit that \emph{exactly} implements $U$, i.e. $U'=U$ (up to some global phase) if and only if the entries of $U$ are in the ring $\intg\left[i,\frac{1}{\sqrt{2}}\right]$ \cite{2013_GS}. We denote this group of unitaries by $\clifft_n$. 
For example, the Toffoli and Fredkin gates belong to $\mathcal{J}_3$. Thus quantum synthesis algorithms can be further sub-divided into two categories : (a) \textbf{exact synthesis algorithms}, that output a circuit implementing $U'=U$ (e.g. \cite{2014_GKMR, 2020_MM}) and (b) \textbf{approximate synthesis algorithms}, that output a circuit implementing $U'$ such that $U'$ is \emph{close to} $U$ (e.g. \cite{2016_RS}).

In this paper we focus on the group $\clifft_n$ of unitaries that can be exactly synthesized and consider the following synthesis problem. \\
\textbf{MIN T-DEPTH} : Given $U\in\clifft_n$ synthesize a T-depth optimal circuit for it. 
In the decision version of this problem we are given $U\in\clifft_n$ and $m\in\nat$, and the goal is to decide if the minimum T-depth of $U$ is at most $m$.

\subsection{Our results}

In this paper we consider the complexity of our exact synthesis algorithms as a function of $m$ and $N=2^n$. We treat arithmetic operations on the entries of $U$ at unit cost, and we do not account for the bit-complexity associated with specifying or manipulating them. 

We first show (in Section \ref{provable:depth}) that the nested meet-in-the-middle (MITM) technique developed in \cite{2020_MM} can be applied to the problem of synthesizing provably depth-optimal circuits. This gives us a depth-optimal-synthesis algorithm with 
time complexity
$O\left(|\mathcal{V}_{n,\mathcal{G}}|^{(c-1)\left\lceil\frac{d'}{c}\right\rceil} \right)$
and space complexity
$O\left(|\mathcal{V}_{n,\mathcal{G}}|^{\left\lceil\frac{d'}{c}\right\rceil} \right)$, where $\mathcal{V}_{n,\mathcal{G}}$ is the set of depth-1 $n$-qubit unitaries over the gate set $\mathcal{G}$, $d'$ is the min-depth of input unitary and $c\geq 2$ is the extent of nesting. This gives us a space-time trade-off for MITM-related techniques applied to this problem. 

Next we apply this technique to synthesize T-depth optimal circuits in Section \ref{provable:Tdepth}. We work with channel representation of unitaries (described in Section \ref{subsec:chanRep}). 
In Section \ref{prelim:Tdepth} we define a ``special'' subset, $\mathbb{V}_n$, of T-depth-1 unitaries, which can generate a T-depth-optimal decomposition of any exactly implementable unitary (up to some Clifford). We prove $|\mathbb{V}_n|\in O(n\cdot 2^{5.6n})$. Then we give an algorithm that returns provably T-depth-optimal circuits and has time and space complexity $O\left(\left(4^{n^2}\right)^{(c-1)\left\lceil\frac{d}{c}\right\rceil}\right)$ and $O\left(\left(4^{n^2}\right)^{\left\lceil\frac{d}{c}\right\rceil} \right)$, respectively, where $d$ is the min-T-depth of input unitary. This is much less than the complexity of the algorithm in \cite{2013_AMMR}. It had a complexity $O(\left(3^n\left|\cliff_n\right|\right)^{\lceil \frac{d}{2}\rceil}\cdot |\cliff_n|)$, where $\cliff_n$ is the set of n-qubit Clifford operators. $|\cliff_n|\in O(2^{kn^2})$ \cite{2008_O, 2014_KS, 1998_CRSS}, for some constant $k>2.5$. In \cite{2013_AMMR} the authors iteratively used $\cliff_n$, as indicated by the stated complexity. It took more than 4 days to generate $\cliff_3$ \cite{2013_AMMR}. 
In fact, in \cite{2013_AMMR} the largest circuit optimally synthesized had 2 qubits and had T-depth 2. We use much smaller sets, which has cardinality $O(4^{n^2})$ and can be derived from $\genv_n$. We can generate $\genv_3$ in a few seconds (Table \ref{tab:Vn}). This gives a (rough) indication about the computational advantage one can have if algorithms are designed with such smaller sets, and thus the motivation to come up with alternate representations.

To improve the efficiency further, we develop another algorithm whose complexity depends on some conjectures that have been motivated by the polynomial complexity algorithm in \cite{2020_MM} for synthesizing T-count optimal circuits. At this point our conjectures do not seem to be derived from the ones in \cite{2020_MM}. If our assumptions are true, then this algorithm returns T-depth-optimal circuits with space and time complexity $\poly(n,2^{5.6n},d)$. Under a weaker assumption this complexity is $\poly(n^{\log n},d,2^{5.6n})$.

 Apart from T-depth-optimal circuit synthesis algorithms for exactly implementable unitaries, the generating set $\mathbb{V}_n$, has found other applications like optimal synthesis algorithms for approximately implementable unitaries \cite{2021_GMM2}.

\subsection{Related work}

The technique of meet-in-the-middle (MITM) and its variant (nested MITM) was used for exact synthesis of provably T-count optimal circuits in \cite{2014_GKMR, 2020_MM} as well as provably depth optimal circuits in \cite{2013_AMMR}. This MITM technique has also been used with deterministic walks in \cite{2016_dMM} to construct a parallel framework for the synthesis of T-count optimal circuits. The time as well as space complexity of the algorithms in \cite{2014_GKMR, 2016_dMM} is $O\left(\left(2^n\right)^m\right)$ where $m$ is the T-count\footnote{The \textbf{T-count of a unitary} is the minimum number of T gates required to implement it.} of the $2^n\times 2^n$ input unitary. The time and space complexity of the algorithm in \cite
{2013_AMMR} is $O\left(\left(3^n\cdot 2^{kn^2}\right)^{\lceil \frac{d}{2}\rceil}\cdot 2^{kn^2}\right)$, where $k$ is a constant and $d$ is the min-T-depth. The first T-count-optimal synthesis algorithm which reduces the complexity to $\poly(2^n,m)$, assuming some conjectures, was given in \cite{2020_MM}. 


 

\subsection{Organization}

We provide some necessary preliminary results and notations in Section \ref{sec:prelim}. Results about T-depth have been derived in Section \ref{prelim:Tdepth}. The algorithms have been given in Section \ref{sec:provable} and \ref{sec:minDepthHeuristic}. Finally we conclude in Section \ref{sec:conclude}.

\section{Preliminaries}
\label{sec:prelim}

 
We write $[K]=\{1,2,\ldots,K\}$. We assume that a set has distinct elements. We denote the $n\times n$ identity matrix by $\id_n$ or $\id$ if the dimension is clear from the context.
The size of an $n$-qubit unitary is denoted by $N=2^n$. 
We call the number of non-zero entries in a matrix as its \textbf{Hamming weight}. We have given some preliminary definitions and facts about the Cliffords, Paulis and the group $\mathcal{J}_n$ generated by Clifford and T gates in Appendix \ref{app:prelim}.
We denote the group of $n$-qubit Pauli operators and Clifford operators by $\pauli_n$ and $\cliff_n$ respectively.
Parenthesized subscripts are used to indicate qubits on which an operator acts. For example, $\X_{(1)}=\X\otimes\id^{\otimes (n-1)}$ implies that Pauli $\X$ matrix acts on the first qubit and the remaining qubits are unchanged.

The \textbf{T-count of a circuit} is the number of T-gates in it. The \textbf{T-count of a unitary} $U$ (denoted by $\tcount(U)$) is the minimum number of T-gates required to implement it. We often simply say, ``T-count'' instead of ``T-count of a unitary''. It should be clear from the context.

\subsection{Channel representations}
\label{subsec:chanRep}

An $n$-qubit unitary $U$ can be completely determined by considering its action on a Pauli $P_s\in\pauli_n$ : $UP_sU^{\dagger}$. 
Since $\pauli_n$ is a basis for the space of all Hermitian $N\times N$ matrices we can write 
\begin{eqnarray}
 UP_sU^{\dagger} = \sum_{P_r\in\pauli_n} \chan{U}_{rs}P_r,\qquad\text{where}\quad
 \chan{U}_{rs} = \frac{1}{2^n}\tr(P_rUP_sU^{\dagger}).  \label{eqn:chanEntry}
\end{eqnarray}
This defines a $N^2\times N^2$ matrix $\chan{U}$ with rows and columns indexed by Paulis $P_r,P_s\in\pauli_n$. We refer to $\chan{U}$ as the \emph{channel representation} of $U$ \cite{2014_GKMR}.

By Hermitian conjugation each entry of the matrix $\chan{U}$ is real. The channel representation respects matrix multiplication, i.e. $\chan{UV}=\chan{U}\chan{V}$. Setting $V=U^{\dagger}$ and using the fact that $\chan{U^{\dagger}}=\big(\chan{U}\big)^{\dagger}$, we see that the channel representation $\chan{U}$ is unitary.
If $U\in\clifft_n$, implying its entries are in the ring $\intg\left[i,\frac{1}{\sqrt{2}}\right]$, then from Equation \ref{eqn:chanEntry} the entries of $\chan{U}$ are in the same ring. Since $\chan{U}$ is real, its entries are from the subring
$$
    \intg\left[\frac{1}{\sqrt{2}}\right]=\left\{\frac{a+b\sqrt{2}}{\sqrt{2}^k} : a,b\in\intg, \quad k\in\nat \right\}
$$
The channel representation identifies unitaries which differ by a global phase. We write the following for the groups in which global phases are modded out.
$$
    \chan{\clifft_n}=\left\{\chan{U}:U\in\clifft_n\right\}, \quad \chan{\cliff_n}=\left\{\chan{C}:C\in\cliff_n   \right\}
$$
Each $Q\in\chan{\cliff_n}$ is a unitary matrix with one nonzero entry in each row and each column, equal to $\pm 1$. This is because Cliffords map Paulis to Paulis up to a possible phase of $-1$. The converse also holds : if $W\in\chan{\clifft_n}$ has this property then $W\in\chan{\cliff_n}$. Since the definition of T-count is insensitive to global phase, it is well-defined in the channel representation and so $\tcount(\chan{U})$ is defined to be equal to $\tcount(U)$.

If a unitary $U$ requires ancilla to be implemented (Theorem \ref{thm:2013_GS}), then we can consider the unitary that acts on the joint state space of input and ancilla qubits. From here on, with a slight abuse of notation when we write $U\in\clifft_n$ we assume it is the unitary that acts on this joint state space. More details can be found in Appendix \ref{app:chan}. However, it is not essential for the rest of the paper.

\begin{definition}
 For any non-zero $v\in\intg\left[\frac{1}{\sqrt{2}}\right]$ the \textbf{smallest denominator exponent}, denoted by $\sde(v)$, is the smallest $k\in\nat$ for which
 $$
    v=\frac{a+b\sqrt{2}}{\sqrt{2}^k} \qquad \text{ with } a,b\in\intg.
 $$
 We define $\sde(0)=0$. \\
 For a $d_1\times d_2$ matrix $M$ with entries over this ring we define 
 $$
    \sde(M)=\max_{a\in[d_1],b\in[d_2]} \sde(M_{ab})
 $$
\end{definition}

\section{T-depth}
\label{prelim:Tdepth}

The purpose of this section is to derive a generating set consisting of T-depth 1 unitaries, such that we can write a T-depth-optimal decomposition of any exactly implementable unitary (up to global phase) as product of elements of this set and a trailing Clifford. This set must be efficiently generated and have a finite cardinality. We first give some essential definitions.

\begin{definition}
The \textbf{depth} of a circuit is the length of any \emph{critical path} through the circuit. Representing a circuit as a directed acyclic graph with nodes corresponding to the circuit's gates and edges corresponding to gate inputs/outputs, a \textbf{critical path} is a path of maximum length flowing from an input of the circuit to an output.
 \label{defn:depth}
\end{definition}
In other words, suppose the unitary $U$ implemented by a circuit is written as a product $U=U_mU_{m-1}\ldots U_1$ such that each $U_i$ can be implemented by a circuit in which all the gates can act in parallel or simultaneously. We say $U_i$ has depth 1 and $m$ is the depth of the circuit. We often refer to each $U_i$ as a \emph{stage} or \emph{(parallel) block}.
The \textbf{T-depth of a circuit} is the number of stages (or unitaries $U_i$) where the $\T/\T^{\dagger}$ gate is the only non-Clifford gate and all the $\T/\T^{\dagger}$ gates can act in parallel. The \textbf{min-T-depth} or \textbf{T-depth of a unitary} $U$ is the minimum T-depth of a Clifford+T circuit that implements it (up to a global phase). We often simply say ``T-depth'' instead of ``T-depth of a unitary''. It should be clear from the context.

Any unitary $U$, having a circuit with T-depth $t$ can be written as follows.
\begin{eqnarray}
 U&=&C_t\left(\tset_{(1)}\ldots\tset_{(n)}\right)C_{t-1}\left(\tset_{(1)}\ldots\tset_{(n)}\right)\ldots C_1\left(\tset_{(1)}\ldots\tset_{(n)}\right)C_0
\end{eqnarray}
In the above equation $\tset\in\{\T,\T^{\dagger},\id\}$ is used to indicate whether there is $\T,\T^{\dagger}$ or $\id$ gate in that qubit. $C_1,C_2,C_3\ldots C_t\in\cliff_n$. For simplicity we ignore the global phase. We can also write the above equation as follows.
\begin{eqnarray}
 U&=&\left(C_t\left(\prod_{i\in [n]}\tset_{(i)}\right)C_t^{\dagger}\right)\left(C_tC_{t-1}\left(\prod_{i\in [n]}\tset_{(i)}\right)(C_tC_{t-1})^{\dagger}\right)\ldots \nonumber\\
 &&\ldots \left(C_tC_{t-1}\ldots C_1\left(\prod_{i\in [n]}\tset_{(i)}\right)(C_tC_{t-1}\ldots C_1)^{\dagger}\right)C_tC_{t-1}\ldots C_1C_0  \nonumber \\
 &=&\left(C_t\left(\prod_{i\in [n]}\tset_{(i)}\right)C_t^{\dagger}\right)\left(C_{t-1}'\left(\prod_{i\in [n]}\tset_{(i)}\right)(C_{t-1}')^{\dagger}\right)\ldots \left(C_1'\left(\prod_{i\in [n]}\tset_{(i)}\right)(C_1')^{\dagger}\right)C_0'\nonumber \\
 &&\qquad\qquad\qquad\qquad [\text{where }C_1',\ldots C_t'\in \cliff_n] \nonumber \\
 &=&V_tV_{t-1}\ldots V_1C_0'\qquad \text{where } V_j=\left(C_j'\left(\prod_{i\in [n]}\tset_{(i)}\right)(C_j')^{\dagger}\right)  \label{eqn:Uinit}
\end{eqnarray}
We call each $V_j$ as a \textbf{(parallel) block}. It is a product of $\T$ or $\T^{\dagger}$ gates on distinct qubits, conjugated by a Clifford. Thus the following set
\begin{eqnarray}
 \genv_n'=\{\prod_{i\in [n]}C\tset_{(i)}C^{\dagger}, C\in\cliff_n, \tset\in\{\T,\T^{\dagger},\id\}\}
 \label{eqn:genv'}
\end{eqnarray}
can be regarded as a \emph{generating set (up to a Clifford) for the decomposition of an exactly implementable unitary}. More precisely, any exactly implementable unitary $U$ (ignoring global phase) can be written as a product of elements from this set and a Clifford. The number of elements from $\genv_n$ is equal to the T-depth of this decomposition or circuit. Any decomposition of $U$ with the minimum number of parallel blocks is called a \textbf{T-depth-optimal decomposition}. A circuit implementing $U$ with the minimum T-depth is called a \textbf{T-depth-optimal circuit}.
 
We can equivalently write each $V_j$ as follows.
\begin{eqnarray}
 V_j=\prod_{i\in [n]}\left(C_j'\tset_{(i)}C_j'^{\dagger}\right)
 \label{eqn:Vj0}
\end{eqnarray}
Now if $C\in\cliff_n$ then 
\begin{eqnarray}
 C\T_{(i)}C^{\dagger}&=&\frac{1}{2}(1+e^{\frac{i\pi}{4}})\id+\frac{1}{2}(1-e^{\frac{i\pi}{4}})CZ_{(i)}C^{\dagger} = \frac{1}{2}(1+e^{\frac{i\pi}{4}})\id+\frac{1}{2}(1-e^{\frac{i\pi}{4}})P \quad [P\in\pm\pauli_n]   \nonumber \\
 &=&R(P) \qquad [\text{Let}]
\end{eqnarray}
The $R(P)$ unitaries and somewhat similar unitaries called \emph{Pauli gadgets} have been studied extensively in previous works like \cite{2014_GKMR, 2019_CDDSS} We believe that the conclusions derived in this paper will enhance the study of these gadgets or special unitaries, such that we can have more applications (for example \cite{2021_GMM2}).

Also $\left(R(P)\right)^{\dagger}=C\T_{(i)}^{\dagger}C^{\dagger}=R^{\dagger}(P)$ (let). Thus we can write Equation \ref{eqn:Vj0} as follows.
\begin{eqnarray}
 V_j=\prod_{i\in [n]}\left(C_j'\tset_{(i)}C_j'^{\dagger}\right)=\prod_{i=n}^1\widetilde{R}(P_{ij}) \qquad [\widetilde{R}\in \{R,R^{\dagger}\}, \widetilde{R}(\id)=\id, P_{ij}\in\pm\pauli_n]
 \label{eqn:Vj}
\end{eqnarray}
The second subscript of $P_{ij}$ gives the index of the block.
The ordering of the intermediate $\T/\T^{\dagger}$ gates does not matter. It merely changes the sequence of $\widetilde{R}( P_{ij})$, but we get the same product $V_j$. Given a set $S$ of qubits there are $3^{|S|}$ possible ways of placing a $\T/\T^{\dagger}/\id$ gate in each qubit. We call each such placement as a \emph{configuration} of $\tset$ gates and denote it by $\tset_{S}$.

From Equation \ref{eqn:Vj} we get a simple way of constructing $\genv_n'$.
\begin{enumerate}
 \item For each $C\in\cliff_n$ do the following. 
 \begin{enumerate}
    \item For each configuration $\tset_{[n]}$ do the following.
    \begin{enumerate}
    \item $V\leftarrow\id$.
     \item For each $i\in [n]$ do the following. \\
        If $\tset_{(i)}\neq\id$ then determine $P=CZ_{(i)}C^{\dagger}$. If $\tset=\T$ then $V\leftarrow V\cdot R(P)$, else  if $\tset=\T^{\dagger}$ then $V\leftarrow V\cdot R^{\dagger}(P)$.
\item Include $V$ in $\genv_n'$ if it does not already exist.
  \end{enumerate}
  \end{enumerate}
  
\end{enumerate}
The time complexity of this procedure is $O\left(|\cliff_n|\right)$ or $O\left(2^{kn^2}\right)$, where $k$ is a constant.
A bound on $|\genv_n'|$ can be obtained by counting all possible distinct $n$-length strings of $\widetilde{R}(P)$, where $\widetilde{R}\in\{R,R^{\dagger}\}$ and $P\in\pm\pauli_n$. Without loss of generality we can assume that every string or sequence is of length $n$, by filling in $R(\id)=R^{\dagger}(\id)=\id$. Thus it gives $|\genv_n'|< \left(2\cdot 2\cdot 4^n\right)^n=4^{n+n^2}$. From \cite{2014_GKMR} we know that there are at most $4^{n^2}\cdot |\cliff_n|$ unitaries (up to global phase) with T-count $n$. So it is highly plausible that $|\genv_n'|\in O(4^{n^2})$. 

Every $n$-length string of $\widetilde{R}(P)$ does not have T-depth 1. We are over-counting a lot here. Our aim is to construct a more compact (smaller) set of T-depth 1 unitaries such that it is possible to write any T-depth 1 unitary as product of unitaries from this set and a Clifford. This is sufficient because it will enable us to write any T-depth-d decomposition (and hence T-depth-optimal decomposition) of a unitary as product of elements from this set and a Clifford (up to global phase). In this way, we can use information from a set of less number of unitaries in order to make more intelligent guesses about a T-depth-optimal decomposition (specially Section \ref{sec:minDepthHeuristic}). We would want to prune many Cliffords to be considered at step 1.  

Here we make the following observation. There are $2^{O(n^2)}$ Clifford operators that can map $Z_{(i)}$ to a particular Pauli $P\in\pauli_n$. All of them lead to the same unitary $R(P)$. Similarly there are many Cliffords such that when $\prod_i Z_{(i)}$ (where the Zs are on different qubits) is conjugated it leads to the same sequence of Paulis (ordering does not matter) i.e. it will give the same unitary $\prod_iR(P_i)$. So for our purpose, what is more important are the \emph{mappings} or rather \emph{images of mappings}, and not the Clifford operators. If $CPC^{\dagger}=P$, we call it a \emph{trivial conjugation}, for any $P\in\pauli_n, C\in\cliff_n$. $P$, in this case, is \emph{trivially conjugated} by $C$.

We now construct a smaller generating set, $\genv_n$. We consider each $\widetilde{R}(P)$ as the starting unit of a string and then determine the remaining $n-1$ units. A formal constructive definition of $\genv_n$ is as follows. 
\begin{definition}
We define $\mathbb{V}_n$, a subset of $n$-qubit unitaries with T-depth 1, that is constructed as follows.
\begin{enumerate}
 \item Include $\widetilde{R}(Z_{(i)})$ ($i\in [n]$) in $\genv_n$. 
 
 \item For each $P\in\pm\pauli_n\setminus\{\id\}$, for each $q\in [n]$ and for each $\widetilde{R}\in\{R,R^{\dagger}\}$ do the following.
\begin{enumerate}
 \item For each Clifford $C$ such that $P=CZ_{(q)}C^{\dagger}$. (If $P=Z_{(q)}$, we will skip this iteration for $Z_{(q)}$. We will discuss later which Cliffords to consider.)
 \begin{enumerate}
  \item For each configuration $\tset_{[n]\setminus\{q\}}$ do the following.
  \begin{enumerate}
   \item $V\leftarrow \widetilde{R}(P)$.
   \item For each $i\in [n]\setminus\{q\}$ do the following.\\
   If $\tset_{(i)}\neq\id$ then determine $P'=CZ_{(i)}C^{\dagger}$. If $\tset=\T$ then $V\leftarrow V\cdot R(P')$, else if $\tset=\T^{\dagger}$ then $V\leftarrow V\cdot R^{\dagger}(P')$.
   \item Include $V$ in $\genv_n$ if it did not already exist.
  \end{enumerate}
 \end{enumerate}
\end{enumerate}
\end{enumerate}

\label{defn:Vn}
\end{definition}
\textbf{Cliffords to be considered (or not considered) at step 2(a) :} We have explained before that for our purpose, combinations of images obtained by conjugating $Z_{(i)}$ (ordering does not matter) is the most important, in order to have distinct unitaries. So we can make some choice of Cliffords to be considered (or rather, not to be considered) at step 2(a). For this, we can make some observations.
\begin{enumerate}
 \item If $C\left(\prod_i\tset_{(i)}\right)C^{\dagger}=\prod_j\widetilde{R}(Z_{(j)})$ for any $C\in\cliff_n$ then it is equal to the unitary $\left(\prod_j\tset_{(j)}\right)$, even if the set of indices $i$ and $j$ are not same. Thus we have included each $\widetilde{R}(Z_{(j)})$ at step 1. Products of these also give T-depth 1 unitaries. In step 2(a) if $P=Z_{(q)}$ then we skip the iteration. In this loop we always consider those sequences of conjugations where there is at least one non-trivial mapping. So we always start with a non-trivial conjugation.
 
 \item If $C=\otimes_iC_i$ for some Cliffords $C_i$ then it is easy to see that we can
 write $U=C\left(\prod_j\tset_{(j)}\right)C^{\dagger}=\prod_iC_i\left(\prod_{j_i}\tset_{(j_i)}\right)C_i^{\dagger}=\prod_i U_i$, where each $U_i$ has T-depth 1. So it is sufficient to consider each $C_i$ and not $C$.
 
 \item Let $U=C\left(\prod_{i=a}^b\tset_{(i)}\right)C^{\dagger}$ is such that $CZ_{(j)}C^{\dagger}=Z_{(j)}$, where $a\leq j\leq b$. Then we can decompose $U=U_1U_2$ where $U_1$ excludes $T_{(j)}$ and $U_2=T_{(j)}$ and each is of T-depth 1. This implies we should be concerned with the images of non-trivial conjugations. (More reason to separate the trivial conjugations at step 1.)  
\end{enumerate}

To determine the Cliffords to be considered we follow the mappings given in \cite{2008_O}. Consider $i\in [n]$. First, we fix $2(4^n-1)4^n$ Cliffords in $\cliff_n$ that conjugate $Z_{(i)}$ or $X_{(i)}$ non-trivially. We call these \emph{coset leaders} of $Z_{(i)}$. The elements of $\cliff_n$ that conjugate $Z_{(i)}$ and $X_{(i)}$ trivially, form a group isomorphic to $\cliff_{n-1}$ with the number of cosets at most $2(4^n-1)4^n$. For example, let $C\in\cliff_n$ is a coset leader (of $Z_{(i)}$) such that $CZ_{(i)}C^{\dagger}=P$ where $P\neq Z_{(i)}$, then any other Clifford that does the same conjugation (which is not a coset leader of $Z_{(i)}$) is of the form $CC'$ where $C'Z_{(i)}C'^{\dagger}=Z_{(i)}$. 
In step 2(a) (when $q=i$) we consider all these coset leaders only. Suppose $C$ is a coset leader that conjugates $Z_{(i)}$ to $P\neq Z_{(i)}$. In the loop 2(a) we considered all possible sequences of $R(P)$ or images obtained by conjugation of $Z_{(j)}$ ($j\neq i$) by $C$. Let $C'$ is non-coset leader of $Z_{(i)}$ and does the trivial conjugation of $Z_{(i)}$. Now among all the $Z_{(j)}$ ($j\neq i$) where $CC'$ conjugates non-trivially, it has to be the coset leader of one of them. This follows from the counting argument. So again we take all possible combinations of images obtained by conjugations by $CC'$, when the loop starts with that particular position of $\T/\T^{\dagger}$. 

\textbf{Taking product :} Now suppose $U_1=C_1\left(\prod_i\tset_{(i)}\right)C_1^{\dagger}\in\genv_n$ and $U_2=C_2\left(\prod_j\tset_{(j)}\right)C_2^{\dagger}\in\genv_n$, and there is no qubit such that a $\T/\T^{\dagger}$-gate is placed in both the unitaries. Let $C_1$ conjugates $Z_{(j)}$ trivially if $j$ is a qubit in which there is a $\T/\T^{\dagger}$ gate in $U_2$. Similarly $C_2Z_{(i)}C_2^{\dagger}=Z_{(i)}$, where there is a $\T/\T^{\dagger}$ gate on qubit $i$ in $U_1$. If $[C_1,C_2]=0$ then it is easy to check that $U=U_1U_2=C_1C_2\left(\prod_k\tset_{(k)}\right)C_2^{\dagger}C_1^{\dagger}$ has T-depth 1. If $C_2Z_{(j)}C_2^{\dagger}=P_j$ and $C_1P_jC_1^{\dagger}=P_j$ then we do not need the commutation condition. It is straightforward to check that these conditions satisfy the 3 observations made earlier\footnote{While constructing $\genv_n$, we can store the information about which unitaries can be multiplied to have a T-depth 1 product.}. Thus we can generate T-depth 1 unitaries (without trailing Clifford) by taking product of unitaries from $\genv_n$.

Thus, from the above discussion we can have the following result.

\begin{theorem}
 Any $U\in\clifft_n$ with T-depth $1$ can be written as follows : $U=e^{i\phi}\left(\prod_{i=d}^1V_i\right)C_0$, where $V_i\in\mathbb{V}_n$, $C_0\in\cliff_n$ and $d\geq 1$.
 \label{claim:dense}
\end{theorem}
\begin{proof}
We ignore the global phase and the trailing Clifford. Let $U=C\left(\prod_{i\in [n]}\tset_{(i)}\right)C^{\dagger}$ (Equation \ref{eqn:Uinit}). Let $S\subseteq [n]$ is the set of qubits such that $C$ conjugates $Z_{(i)}$ trivially, where $i\in S$. Then we can write $U=\left(\prod_{i\in S}\widetilde{R}(Z_i)\right)C\left(\prod_{i\in \overline{S}}\tset_{(i)}\right)C^{\dagger}=\left(\prod_{i\in S}\widetilde{R}(Z_i)\right)U'.$ Each of these $\widetilde{R}(Z_{(i)})$ are included in $\genv_n$ (step 1). So now let us consider the second term, $U'$, in the product. If $C=\otimes_{j}C_j$ then we can write $U'=\prod_jC_j\left(\prod_{k\in S_j}\tset_{(k)}\right)C_j^{\dagger}=\prod_jU_j'$, where $S_j\subseteq \overline{S}$ is the set of qubits on which $C_j$ acts. If there are no $\T/\T^{\dagger}$ gates at any qubit of $S_j$ then $C_jC_j^{\dagger}=\id$. Else, there exists at least one $k\in S_j$ such that $C_j$ conjugates $Z_{(k)}$ non-trivially. In step 2 of the definition of $\genv_n$, we have included each such $U_j$ in our set. This proves the theorem.
\end{proof}

In \cite{2014_GKMR} it has been shown that $\{R(P):P\in\pauli_n\}$ generates the T-count-optimal decomposition of any exactly implementable unitary, up to a Clifford. The channel representation inherits these decompositions and in this representation the global phase goes away. Thus we can write the following.
\begin{eqnarray}
 \chan{U}=\Big(\prod_{i=d}^{1} \chan{V_i} \Big)\chan{C_0}
 \label{eqn:decomposeChan}
\end{eqnarray}
Let
\begin{eqnarray}
 \chan{\mathbb{V}_n} = \{\chan{V'} : V'\in\mathbb{V}_n \}.
 \label{eqn:VnChan}
\end{eqnarray}

\begin{fact}
$
    |\mathbb{V}_n| \leq 2n\cdot 3^{n-1}\cdot 4^n\cdot 4^n < n\cdot 2^{5.6n}
$ and hence
$
    |\chan{\mathbb{V}_n}| < n\cdot 2^{5.6n}
$.
 \label{fact:Vn}
\end{fact}
\begin{proof}
From Definition \ref{defn:Vn}, for each starting $R(P)/R^{\dagger}(P)$ there can be $n$ positions for first $\T/\T^{\dagger}$ gate respectively. In the remaining qubits we can have $\T,\T^{\dagger}$ or $\id$. Thus there are at most $3^{n-1}$ ways to place the $\T/\T^{\dagger}$ gates in remaining $(n-1)$ qubits. Given a starting Clifford and a configuration, the rest of the $R(P)$ unitaries are uniquely determined. We have discussed that we need to consider at most $2\cdot 4^n \cdot 4^n$ Cliffords (coset leaders, as discussed before) that can map each $Z_{(i)}$ to any $P$ \cite{2008_O, 2014_KS, 1998_CRSS}. More precisely, there are at most $2\cdot 4^n\cdot 4^n$ choices for the starting Clifford for each of the $n$ positions of the starting $\T/\T^{\dagger}$ gate, that can lead to distinct strings of $R(P)$ during the construction of $\genv_n$ . So we get the stated bounds.
\end{proof}
In Table \ref{tab:Vn} we have compared the cardinalities and generation time of $\genv_n$ and $\cliff_n$. The latter has been used in \cite{2013_AMMR} to design a T-depth-optimal-synthesis algorithm. We use the set $\genv_n$ for our heuristic algorithm in Section \ref{sec:minDepthHeuristic}. In the next section we use a bigger set with cardinality $O(4^{n^2})$, much less than $|\cliff_n|\in O(2^{kn^2})$, where $k>2.5$. This set can be derived from $\genv_n$, or we can simply use $\genv_n'$. We will see in the following sections how the cardinalities of these sets make a difference in the running time and space of the various algorithms. 
\begin{table}[!htbp]
\centering
\begin{tabular}{|c|c|c|c|c|}
 \hline
 $\#$Qubits (n) & $\left|\mathbb{V}_n\right|$ & Generation time & $\left|\cliff_n\right|$ &  Generation time \cite{2013_AMMR}\\
 \hline
 2 & $122$ & 0.015s & $\approx 11520 $ & 1s \\
 \hline
 3 & $2282$ & 2.212s & $\approx 92,897,280$ & $> $ 4 days\\
 \hline 
 4 & $35846$ & 10m 24s & N/A & N/A \\
 \hline
 \end{tabular}
 \caption{Comparison of generation time of $\genv_n$ and $\cliff_n$.}
 \label{tab:Vn}
 \end{table}

The following fact can be easily proved from Fact 3.2 in \cite{2020_MM}.
\begin{fact}[\cite{2020_MM}]
Let $W'=\chan{\widetilde{R}(P)}W$ where $W$ and $W'$ are unitaries, $\widetilde{R}\in\{R,R^{\dagger}\}$ and $P\in\pm\pauli_n$. Then $\sde(W')=\sde(W)\pm 1$ or $\sde(W')=\sde(W)$.
 \label{fact:sdeChange}
\end{fact}
Some more information about the properties of $\chan{R(P)}$ and $\chan{R^{\dagger}(P)}$ have been provided in Appendix \ref{app:RP}. This will help in computing $\chan{\genv_n}$ faster, but it will not make much difference in the asymptotic complexity of any of our algorithms. So these are not essential for the rest of the paper.

\section{A faster synthesis algorithm for T-depth}
\label{sec:provable}

In this section we describe an exact synthesis algorithm that finds a circuit that is provably T-depth-optimal. We modify the algorithm by Amy et al. \cite{2013_AMMR} and employ a nested meet-in-the-middle technique, as has been done by Mosca and Mukhopadhyay \cite{2020_MM}, to optimize T-count. This gives more space efficient algorithm to get optimal depth circuit. Furthermore, we work with channel representations to get T-depth-optimal circuits. This reduces both the time and space complexity compared to the algorithm in \cite{2013_AMMR}.

\subsection{An exact algorithm for depth-optimal circuits}
\label{provable:depth}

We first describe a general algorithm where we are given a set of gates (and their inverses), $\mathcal{G}$, with which we want to design a depth optimal circuit implementing a unitary $U$. The set $\mathcal{G}$ is called the \emph{instruction set}. Let $\mathcal{V}_{n,\mathcal{G}}$ be the set of $n$-qubit unitaries of depth 1 that can be implemented by a circuit designed with the gates in $\mathcal{G}$.
We state the following lemma which can be regarded as a generalization of Lemma 1 in \cite{2013_AMMR}. The proof has been given in Appendix \ref{sec:appendix}. This observation allows us to search for circuits of depth $d$ by only generating circuits of depth at most $\left\lceil\frac{d}{c}\right\rceil$ ($c\geq 2$).

\begin{lemma}
Let $S_i\subset U(2^n)$ be the set of all unitaries implementable in depth $i$ over the gate set $\mathcal{G}$. Given a unitary $U$, there exists a circuit over $\mathcal{G}$ of depth $(d_1+d_2)$ implementing $U$ if and only if $S_{d_1}^{\dagger}U\bigcap S_{d_2}\neq\emptyset$.
 \label{lem:provNest}
\end{lemma}
We now describe our procedure (Nested MITM), whose pseudocode has been given in Appendix \ref{sec:app:code} (Algorithm \ref{app:alg:nestMITM}). The input consists of the unitary $U$, instruction set $\mathcal{G}$, depth $d$ and $c\geq 2$ that indicates the extent of nesting or recursion we want in our meet-in-the-middle approach. If $U$ is of depth at most $d$ then the output consists of a decomposition of $U$ into smaller depth unitaries, else the algorithm indicates that $U$ has depth more than $d$. At the beginning of the algorithm 
we generate the set $\mathcal{V}_{n,\mathcal{G}}$.

The algorithm consists of $\left\lceil\frac{d}{c}\right\rceil$ iterations 
and in the $i^{th}$ such iteration we generate circuits of depth $i$ ($S_{i}$) by extending the circuits of depth $i-1$ ($S_{i-1}$) by one more level. Then we use these two sets to search for circuits of depth at most $ci$ 
. The search is performed iteratively where in the $k^{th}$ ($1\leq k\leq c-1$) round we generate unitaries of depth at most $ki$ by taking $k$ unitaries $W_1,W_2,\ldots,W_k$ where $W_i\in S_i$ or $W_i\in S_{i-1}$. Let $W=W_1W_2\ldots W_k$ and its depth is $k'\leq ki$. We search for a unitary $W'$ in $S_i$ or $S_{i-1}$ such that $W^{\dagger}U=W'$. By Lemma \ref{lem:provNest} if we find such a unitary it would imply that depth of $U$ is $k'+i$ or $k'+i-1$ respectively. In the other direction if the depth of $U$ is either $k'+i$ or $k'+i-1$ then there should exist such a unitary $W'$ in $S_i$ or $S_{i-1}$ respectively. Thus if the depth of $U$ is at most $d$ then the algorithm terminates in one such iteration and returns a decomposition of $U$. This proves the \textbf{correctness} of this algorithm.

\paragraph{Time and space complexity}

We impose a strict lexicographic ordering on unitaries such that a set $S_i$ can be sorted with respect to this ordering in $O\left(|S_i|\log |S_i|\right)$ time and we can search for an element in this set in $O\left(\log |S_i|\right)$ time. An example of such an ordering is ordering two unitaries according to the first element in which they differ. Now consider the $k^{th}$ round of the $i^{th}$ iteration (steps \ref{nestMITM:whileStart}-\ref{nestMITM:whileEnd} of Algorithm \ref{app:alg:nestMITM} in Appendix \ref{sec:app:code}).
We build unitaries $W$ of depth at most $ki$ using elements from $S_i$ or $S_{i-1}$. Number of such unitaries is at most $|S_i|^{k}$. Given a $W$, time taken to search for $W'$ in $S_i$ or $S_{i-1}$ such that $W^{\dagger}U=W'$ is $O\left(\log |S_i|\right)$. Since $|S_j|\leq |\mathcal{V}_{n,\mathcal{G}}|^j$, so the $k^{th}$ iteration of the for loop within the $i^{th}$ iteration of the while loop, takes time $O\left(|\mathcal{V}_{n,\mathcal{G}}|^{(c-1)i}\log |\mathcal{V}_{n,\mathcal{G}}| \right)$. Thus the time taken by the algorithm is $O\left(|\mathcal{V}_{n,\mathcal{G}}|^{(c-1)\left\lceil\frac{d}{c}\right\rceil} \log |\mathcal{V}_{n,\mathcal{G}}| \right)$. 

In the algorithm we store unitaries of depth at most $\left\lceil\frac{d}{c}\right\rceil$. So the space complexity of the algorithm is $O\left(|\mathcal{V}_{n,\mathcal{G}}|^{\left\lceil\frac{d}{c}\right\rceil} \right)$.
Since $|\mathcal{V}_{n,\mathcal{G}}|\in O\left(|\mathcal{G}|^n \right)$, so we have an algorithm with space complexity $O\left(|\mathcal{G}|^{n\left\lceil\frac{d}{c}\right\rceil} \right)$ and time complexity $O\left(n|\mathcal{G}|^{n(c-1)\left\lceil\frac{d}{c}\right\rceil} \log |\mathcal{G}| \right)$.
\subsection{Reducing both space and time complexity to find T-depth optimal circuits}
\label{provable:Tdepth}

We now consider the special case where $\mathcal{G}$ is the Clifford+T gate set and the goal is to design a T-depth optimal circuit for a given unitary $U$. We work with the channel representation of unitaries. We generate the set $\genv_n''$, which consists of products of unitaries from $\genv_n$ and has T-depth 1. We have explained in Section \ref{prelim:Tdepth} how to perform such products. We can even use $\genv_n'$ described in the previous section. In Section \ref{prelim:Tdepth} we gave conditions for generating these products. Thus we replace $\mathcal{V}_{n,\mathcal{G}}$ with $\chan{\mathbb{V}_n''}$. 
It is easy to see that for any T-depth 1 unitary $\chan{U}$ there exists $\chan{V}\in\chan{\mathbb{V}_n''}$ such that $\chan{U}=\chan{V}\chan{C}$ for some Clifford $C\in\cliff_n$. This motivates us to use the following definition from \cite{2014_GKMR}.
\begin{definition}[\textbf{Coset label}]
Let $W\in\chan{\clifft_n}$. Its coset label $W^{(co)}$ is the matrix obtained by the following procedure.
(1) Rewrite $W$ so that each nonzero entry has a common denominator, equal to $\sqrt{2}^{\sde(W)}$. (2) For each column of $W$, look at the first non-zero entry (from top to bottom) which we write as $v=\frac{a+b\sqrt{2}}{\sqrt{2}^{\sde(W)}}$. If $a<0$, or if $a=0$ and $b<0$, multiply every element of the column by $-1$. Otherwise, if $a>0$, or $a=0$ and $b>0$, do nothing and move on to the next column. (3) After performing this step on all columns, permute the columns so that they are ordered lexicographically from left to right.

 \label{defn:cosetLabel}
\end{definition}
Since unitaries stored in $\chan{\mathbb{V}_n''}$ are distinct, we can say that this set stores the coset labels of T-depth 1 unitaries. The following can be shown.
\begin{theorem}[\textbf{Proposition 2 in \cite{2014_GKMR}}]
Let $W,V\in\chan{\clifft_n}$. Then $W^{(co)}=V^{(co)}$ if and only if $W=VC$ for some $C\in\chan{\cliff_n}$.
 \label{thm:cosetLabel}
\end{theorem}

The nested meet-in-the-middle search for T-depth-optimal circuit is performed as described before, except for the following changes. We replace the set $\mathcal{V}_{n,\mathcal{G}}$ with the set $\chan{\mathbb{V}_n''}$ (step \ref{nestMITM:depthInc} of Algorithm \ref{app:alg:nestMITM}), for reasons described before. This helps us to generate coset labels of unitaries with increasing T-depth. We work with channel representations $\chan{W},\chan{W_1},\chan{W_2},\ldots$. So at the $k^{th}$ round of the $i^{th}$ iteration, we calculate $\chan{W}=\prod_{j=1}^k\chan{W_j}$ where $\chan{W_j}\in S_i$ or $S_{i-1}$. Then we check if $\exists \chan{W'}\in S_i$ (or $S_{i-1}$ respectively) such that $\left(\chan{W}^{\dagger}\chan{U}\right)^{(co)}=\chan{W'}$. If such a unitary exists it would imply $U=e^{i\phi}WW'C$ for some Clifford $C\in\cliff_n$. From Lemma \ref{lem:provNest} we can say that $U$ can be implemented by a circuit with T-depth equal to the sum of the T-depth of the circuit for $W$ and $W'$. 

\paragraph{Space and time complexity}
From Fact \ref{fact:Vn} we know that $|\chan{\mathbb{V}_n}|\leq n\cdot 2^{5.6n}$ and $\genv_n''$ is formed by taking product of unitaries from $\genv_n$. The most naive upper bound that we can have is $|\genv_n''|\in O(4^{n^2})$, which is the bound on $\genv_n'$ discussed in Section \ref{prelim:Tdepth}.\footnote{We believe that $|\genv_n''|$ is much less than $4^{n^2}$.}
Thus analysing in the same way as before we can say that the algorithm has space complexity $O\left((4^{n^2})^{\lceil\frac{d}{c}\rceil} \right)$ and time complexity $O\left((4^{n^2})^{(c-1)\lceil\frac{d}{c}\rceil} \right)$ ($c\geq 2$). This is much less than the space and time complexity of the T-depth-optimal algorithm in \cite{2013_AMMR}. They use the MITM technique and the space and time complexity is $O\left(\left(3^n|\cliff_n|\right)^{\lceil \frac{d}{2}\rceil}\cdot |\cliff_n|\right)$. The cardinality of the n-qubit Clifford group, $\cliff_n$, is $O(2^{kn^2})$ ($k>2.5$) \cite{2008_O, 2014_KS}. So the space and time complexity is $O\left(\left(2^{kn^2}\right)^{\lceil\frac{d}{2}\rceil+1}3^{n\lceil\frac{d}{2}\rceil}\right)$, where $k>2.5$.  Clearly, even if the extent of nesting is 2 i.e. $c=2$, in which case our procedure becomes a MITM algorithm, we get a significant improvement in both time and space complexity.

\section{A more efficient algorithm to synthesize T-depth optimal circuits}
\label{sec:minDepthHeuristic}

In this section we describe an algorithm that on input a $2^n\times 2^n$ unitary $U$ finds a T-depth optimal circuit for it and has space and time complexity $\poly(n, 2^{5.6n},d)$ with some conjecture (or $\poly(n^{\log n},d,2^{5.6n})$ with a weaker conjecture), where $d$ is the min-T-depth of $U$. We draw inspiration from some observations made in \cite{2020_MM}, while developing a polynomial time algorithm for synthesizing T-count-optimal circuits. We came up with another novel way of pruning the search space. The numerical results of this section (Table~\ref{tab:benchmark} and Table~\ref{tab:random}) are available online at \url{https://github.com/vsoftco/t-depth}.

The input of our algorithm is the channel representation of a $2^n\times 2^n$ unitary $U$. From Theorem \ref{claim:dense} (Section \ref{prelim:Tdepth}) we know there exists a T-depth-optimal decomposition of $\chan{U}$ as follows : $\chan{U}=\left(\prod_{i=d''}^1 \chan{V_i} \right)\chan{C_0}$, where $C_0\in\cliff_n$, $\chan{V_i}\in\mathbb{V}_n$, $d\leq d''\leq dn$ and $d$ is the T-depth of $U$. We iteratively try to guess the blocks $\chan{V_i}$ by looking at the change in some ``properties'' of the matrix $\chan{V_i}^{-1}\chan{U'}$ where $\chan{U'}=\prod_{j=d''}^{i+1}\chan{V_j}^{-1}\chan{U}$. If we have the correct sequence then we should reach $\chan{C_0}$, a matrix consisting of exactly one $+1$ or $-1$ in each row and column. As in \cite{2020_MM} we consider two properties of the resultant matrices - their sde and Hamming weight. The intuition is as follows. Consider a unitary $\chan{W}$ and we multiply it by $\chan{V_1}\in\chan{\mathbb{V}_n}$. Let $\chan{Y}=\chan{WV_1}$, $\Delta_s=\sde(\chan{W})-\sde(\chan{Y})$ and $\Delta_h=\ham(\chan{W})-\ham(\chan{Y})$, where $\ham(.)$ is the Hamming weight. Now we multiply $\chan{Y}$ by $\chan{V_i}^{-1}$ where $\chan{V_i}\in\chan{\mathbb{V}_n}$. Let $\chan{Z}=\chan{YV_i^{-1}}$, $\Delta_s^i=\sde(\chan{Y})-\sde(\chan{Z})$ and $\Delta_h^i=\ham(\chan{Y})-\ham(\chan{Z})$. If $V_i=V_1$ then $\Delta_s=-\Delta_s^i$ and $\Delta_h=-\Delta_h^i$. But if $V_i\neq V_1$ then with high probability we do not expect to see this kind of change. This helps us to distinguish the $V_i$'s in at least one T-depth-optimal decomposition.

The pseudocode for algorithm MIN T-DEPTH has been given in Appendix \ref{sec:app:code} (Algorithm \ref{app:alg:minTdepth}). We iteratively call the sub-procedure $\mathcal{A}(\chan{U},d')$ with the value $d'\in\intg$ increasing in each iteration. We accumulate all decompositions returned by $\mathcal{A}$. Then in MIN T-DEPTH we check if in each such decomposition we can combine consecutive unitaries to form a T-depth 1 unitary (Section \ref{prelim:Tdepth}). We output a decomposition with the minimum T-depth. Here let us explain the starting value for $d'$.
If we know that any circuit requires at least $x$ T gates to implement $U$, we know that the T-depth of any circuit implementing $U$ will be at least $\lceil\frac{x}{n}\rceil$. Thus if we know $\tcount(U)$ i.e. the T-count of $U$ we can start the iterations with $d'=\lceil\frac{\tcount(U)}{n}\rceil$. If we do not know that, we can consider $\sde(\chan{U})$. Due to Fact \ref{fact:sdeChange} we know $\tcount(U)\geq\sde(\chan{U})$, so we can also start the iterations with $d'=\lceil\frac{\sde(\chan{U})}{n}\rceil$. We can also determine stopping criteria from these information. For example, if we get a decomposition with T-depth $\lceil\frac{\tcount(U)}{n}\rceil$, then we can stop immediately. Alternatively, we can generate the set $\genv_n''$, described in Section \ref{sec:provable} and stop as soon as we get a decomposition in $\mathcal{A}$. 

\begin{figure}[h]
\begin{tikzpicture}
 \draw [ultra thick] (7.5,0) circle [radius=0.3];
 \node at (7.5,0) {$\chan{U}$};
 
 \draw [thick] (0,-1.5) rectangle (3.4,-2);
 \draw [fill=black] (0.3,-1.75) circle [radius=0.25];
 \draw [fill=black] (1,-1.75) circle [radius=0.25];
 \draw [fill=black] (1.5,-1.75) circle [radius=0.05];
 \draw [fill=black] (1.75,-1.75) circle [radius=0.05];
 \draw [fill=black] (2,-1.75) circle [radius=0.05];
 \draw [thick] (2.5,-1.75) circle [radius=0.25];
 \draw (2.8,-1.5)--(2.8,-2);    
 \node at (3.1,-1.75) {$1$};
 \draw (7.2,-0.1)--(0,-1.5);    
 \draw (7.2,-0.1)--(2.5,-1.5);
 \node at (2.85,-1.1) {$1$};
 
 \draw [thick] (3.8,-1.5) rectangle (7.2,-2);
 \draw (6.6,-1.5)--(6.6,-2);
 \node at (6.9,-1.75) {$2$};
 \draw (7.25,-0.11)--(3.8,-1.5);    
 \draw (7.3,-0.16)--(6.3,-1.5);
 \node at (6,-1.1) {$2$};
 \draw [thick] (4.1,-1.75) circle [radius=0.25];    
 \draw [fill=black] (4.8,-1.75) circle [radius=0.25];
 \draw [fill=black] (5.3,-1.75) circle [radius=0.05];
 \draw [fill=black] (5.55,-1.75) circle [radius=0.05];
 \draw [fill=black] (5.8,-1.75) circle [radius=0.05];
 \draw [thick] (6.3,-1.75) circle [radius=0.25];
 
 \draw [fill=black] (8,-1.75) circle [radius=0.05]; 
 \draw [fill=black] (8.5,-1.75) circle [radius=0.05];
 \draw [fill=black] (9,-1.75) circle [radius=0.05];
 \draw [fill=black] (9.5,-1.75) circle [radius=0.05];
 \draw [fill=black] (10,-1.75) circle [radius=0.05];
 
 \draw [thick] (11,-1.5) rectangle (14.4,-2);
 \draw (13.8,-1.5)--(13.8,-2);
 \node at (14.1,-1.75) {$n$};
 \draw (7.8,-0.05)--(11.2,-1.5);  
 \draw (7.8,-0.05)--(13.7,-1.5);
 \node at (11.2,-1.1) {$n$};
 \draw [fill=black] (11.3,-1.75) circle [radius=0.25];
 \draw [thick] (12,-1.75) circle [radius=0.25];
 \draw [fill=black] (12.5,-1.75) circle [radius=0.05];
 \draw [fill=black] (12.75,-1.75) circle [radius=0.05];
 \draw [fill=black] (13,-1.75) circle [radius=0.05];
 \draw [thick] (13.5,-1.75) circle [radius=0.25];

 \draw [thick] (0,-3) rectangle (3.2,-3.5);
 \draw (2.8,-3)--(2.8,-3.5);
 \node at (3,-3.25) {$2$};
 \draw (0.3,-2)--(0.3,-3);  
 \draw (1,-2)--(2.6,-3);
 \node at (1.1,-2.8) {$1$};
 \draw (0.3,-2)--(3.6,-3);  
 \draw (1,-2)--(5.9,-3);
 \node at (2.7,-2.5) {$2$};
 \draw [thick] (0.3,-3.25) circle [radius=0.25];
 \draw [thick] (1,-3.25) circle [radius=0.25];
 \draw [fill=black] (1.5,-3.25) circle [radius=0.05];
 \draw [fill=black] (1.75,-3.25) circle [radius=0.05];
 \draw [fill=black] (2,-3.25) circle [radius=0.05];
 \draw [thick] (2.5,-3.25) circle [radius=0.25];
 
 \draw [thick] (3.4,-3) rectangle (6.6,-3.5);
 \draw (6.2,-3)--(6.2,-3.5);
 \node at (6.4,-3.25) {$3$};
 \draw (4.8,-2)--(3.6,-3);    
 \draw (4.8,-2)--(5.9,-3);
 \node at (4.8,-2.5) {$1$};
 \draw (4.8,-2)--(7.1,-3);  
 \draw (4.8,-2)--(9.3,-3);
 \node at (6.2,-2.5) {$2$};
 \draw [fill=black] (3.7,-3.25) circle [radius=0.25];
 \draw [thick] (4.4,-3.25) circle [radius=0.25];
 \draw [fill=black] (4.9,-3.25) circle [radius=0.05];
 \draw [fill=black] (5.15,-3.25) circle [radius=0.05];
 \draw [fill=black] (5.4,-3.25) circle [radius=0.05];
 \draw [thick] (5.9,-3.25) circle [radius=0.25];
 
 \draw [thick] (6.8,-3) rectangle (10,-3.5);
 \draw (9.6,-3)--(9.6,-3.5);
 \node at (9.8,-3.25) {$4$};
 \draw [thick] (7.1,-3.25) circle [radius=0.25];
 \draw [thick] (7.8,-3.25) circle [radius=0.25];
 \draw [fill=black] (8.3,-3.25) circle [radius=0.05];
 \draw [fill=black] (8.55,-3.25) circle [radius=0.05];
 \draw [fill=black] (8.8,-3.25) circle [radius=0.05];
 \draw [thick] (9.3,-3.25) circle [radius=0.25];
 
 \draw [fill=black] (10.4,-3.25) circle [radius=0.05];  
 \draw [fill=black] (10.9,-3.25) circle [radius=0.05];
 \draw [fill=black] (11.4,-3.25) circle [radius=0.05];
 
 \draw [thick] (11.8,-3) rectangle (15,-3.5);
 \draw (14.6,-3)--(14.6,-3.5);
 \node at (14.8,-3.25) {$2n$};
 \draw (11.3,-2)--(12,-3);    
 \draw (11.3,-2)--(14.3,-3);
 \node at (12,-2.5) {$n$};
 \draw [thick] (12.1,-3.25) circle [radius=0.25];
 \draw [thick] (12.8,-3.25) circle [radius=0.25];
 \draw [fill=black] (13.3,-3.25) circle [radius=0.05];
 \draw [fill=black] (13.55,-3.25) circle [radius=0.05];
 \draw [fill=black] (13.8,-3.25) circle [radius=0.05];
 \draw [thick] (14.3,-3.25) circle [radius=0.25];
\end{tikzpicture}
\caption{The tree built in $\mathcal{A}$ (Algorithm \ref{app:alg:Aheuristic}). Each node stores a unitary, the root at level 0 storing $\chan{U}$. The edges are labeled by unitaries in $\chan{\mathbb{V}_n}^{-1}$. A child node unitary is obtained by multiplying the edge unitary with the parent node unitary. The edges are grouped into hyper-edges, where each hyper-edge is labeled by a unitary in $\chan{\mathbb{V}_{n,j}}^{-1}$. The nodes are grouped into hyper-nodes, where each hyper-node has a number indicating the number of $\chan{R(P)}^{-1}$ in the path from the root to each node in this hyper-node. Within each hyper-node we select some nodes according to some criteria and the nodes in the next level are built from these selected (black) nodes.  }
 \label{fig:minDepthHeuristic}
\end{figure}
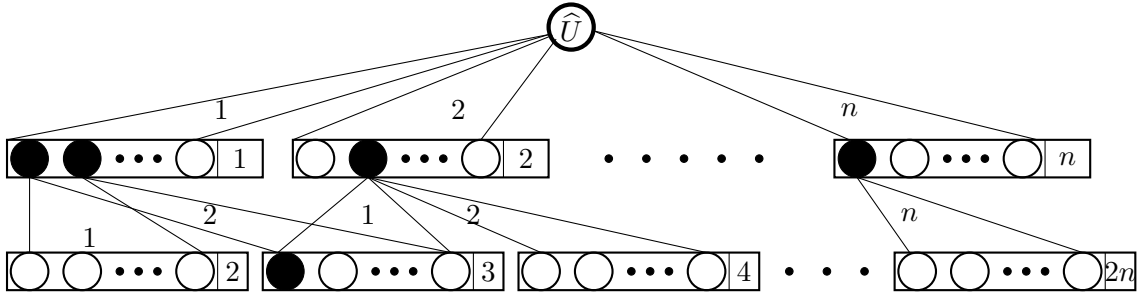

It will be useful if we depict the procedure $\mathcal{A}$ using a tree (Figure \ref{fig:minDepthHeuristic}), where each node stores a unitary.
The root (depth $0$) stores $\chan{U}$. The edges are labeled by unitaries from $\chan{\mathbb{V}_n}^{-1}$, which is defined as
$$
    \chan{\mathbb{V}_n}^{-1}=\{\chan{V}^{-1}:\chan{V}\in\chan{\mathbb{V}_n} \}
$$
This is a set of $n$-qubit unitaries with T-depth $1$ (refer Section \ref{prelim:Tdepth}). A child node unitary is obtained by multiplying the parent unitary with the unitary of the edge. We refer to these two types of unitaries as ``node-unitary'' and ``edge-unitary'' respectively. The product of the edge unitaries on a path from the root to a non-root node is referred to as the ``path unitary'' with respect to the non-root node. By ``path T-count'' of a non-root node we refer to the sum of the number of $R(P)$ terms in the edge-unitaries. Each $R(P)$ has one T-gate. At each depth of the tree we group the nodes into some ``hypernodes'' such that the path T-count of each node within a hypernode is same. At this point it will be useful to observe $\chan{\mathbb{V}_n}^{-1}=\bigcup_{1\leq j\leq n} \chan{\mathbb{V}_{n,j}}^{-1}$, where $\chan{\mathbb{V}_{n,j}}^{-1}$ is the set of unitaries with $j$ number of $\chan{R(P)}^{-1}$. 
In Figure \ref{fig:minDepthHeuristic} we have grouped the edges such that the edge-unitaries within one such ``hyperedge'' are from $\chan{\mathbb{V}_{n,j}}^{-1}$ for some $j$.

At each depth, within each such hypernode we sub-divide the nodes according to the sde of its unitary and change in Hamming weight of this unitary compared to the parent node-unitary. By change in Hamming weight we mean if it has increased or decreased or remains unchanged, with respect to the Hamming weight of the parent node. Within each hypernode we select the set of nodes with minimum cardinality such that sde of its unitaries can be reduced to $0$ within depth $d'$ of the tree. We build the nodes in the next level from the ``selected'' node-unitaries only. We stop building the tree as soon as we reach a node-unitary with sde $0$, indicating we reached a Clifford. If we have not reached any Clifford within depth $d'$ we quit and conclude that minimum T-depth of $\chan{U}$ is more than $d'$. A pseudocode of the procedure $\mathcal{A}$ has been given in Appendix \ref{sec:app:code} (Algorithm \ref{app:alg:Aheuristic}). The number of hypernodes in depth $i$ can be at most $ni-i+1$, since the path T-count of any unitary can be at most $ni$ and at least $i$. Also, since the sde can change by at most $1$ after multiplying by any $\chan{R(P)}^{-1}$ (Fact \ref{fact:sdeChange}), then after multiplying by any unitary in $\mathbb{V}_{n,j}^{-1}$ sde of any unitary can change by at most $j$. So (at step \ref{Aheuristic:takeMin} of Algorithm \ref{app:alg:Aheuristic}) we select the minimum sized set among those sets of unitaries which has the potential to reach the Clifford within the remaining steps.

To analyse the space and time complexity of our algorithm we make the following conjecture.
\begin{conjecture}
(a) While dividing the nodes according to their sde and change in Hamming weight within any hypernode, the minimum cardinality of any set (such that its sde can be potentially reduced to $0$) is bounded by $\poly(2^n)$. (b) Also, we get at least one T-depth-optimal decomposition.
 \label{conj:minDepth}
\end{conjecture}
 So our conjecture has two parts. (a) bounds the size of the tree and thus determines the complexity of the algorithm. (b) implies that we can preserve at least one T-depth-optimal decomposition by pruning in this way. So it determines the efficiency. We can make a weaker conjecture with a more relaxed bound.
\begin{conjecture}[\textbf{Weaker version}]
(b) While dividing the nodes according to their sde and change in Hamming weight within any hypernode, the minimum cardinality of any set (such that its sde can be potentially reduced to $0$) is bounded by $\poly(n^{\log n},2^n)$. (b) Also, we get at least one T-depth-optimal decomposition. 
\label{conj:minDepth2}
\end{conjecture}

\paragraph{Comparison with Conjecture 1 in \cite{2020_MM}}

In \cite{2020_MM} the authors proposed some conjectures to reduce the complexity of synthesizing T-count-optimal circuits. Our algorithm has been motivated by that work but based on current knowledge it does not appear that Conjectures 1 or 2 can be derived from the conjecture used in \cite{2020_MM}, with the present knowledge. The main intuition of these conjectures stem from the following observation. Suppose we multiply a unitary $\chan{U}'$ by $\chan{R(P_1)}$. We will notice some change in the properties (like sde, Hamming weight) in the product unitary $\chan{W}'=\chan{U}'\chan{R(P_1)}$ compared to the initial $\chan{U}'$. Now when we multiply $\chan{W}'$ by $\chan{R(P_1)}^{-1}$ we will see these effects reversed. But if we multiply $\chan{W}'$ by $\chan{R(P_i)}^{-1}$ (where $i\neq 1$) then with high probability we will observe some other effects. In \cite{2020_MM} the authors used these intuitions to design a T-count-optimal algorithm, where they iteratively tried to guess a sequence of $R(P)$s in a T-count-optimal decomposition of $U$, by observing these change in properties. 
In our present algorithm (see Figure \ref{fig:minDepthHeuristic}) we consider many paths with different T-counts at each level. Now the T-depth-optimal decompositions will follow some of these paths. 
When we select the minimum cardinality set in each hypernode (where all unitaries have same path T-count), we expect that the ``distinguishing property'' that we explained before does not get destroyed even if we multiply an (intermediate) unitary by up to $n$ $\chan{R(P)}^{-1}$. We do not see how this observation follows from the conjecture in \cite{2020_MM}, without some more knowledge about the underlying mathematics. So for T-depth-optimal decompositions we have made separate conjectures.

\paragraph{Space and time complexity}

We consider the time and space complexity of $\mathcal{A}$.
From Fact \ref{fact:Vn} we know $|\chan{\mathbb{V}_n}^{-1}|\leq n\cdot 2^{5.6n}$. These are the number of unitaries we always store. 

In the $i^{th}$ iteration we have up to $ni-i+1$ children hypernodes. There are at most $n(i-1)-(i-1)+1$ parent hypernodes and within each at most $\poly(2^n)$ parent nodes are selected by Conjecture \ref{conj:minDepth}. Each parent node is multiplied by $|\chan{\mathbb{V}_n}|$ $2^{2n}\times 2^{2n}$ unitaries. Arguing in similar way space and time complexity of procedure $\mathcal{A}$ is $\poly\left(d',n, 2^{5.6n}\right)$.

Since MIN T-DEPTH consists of at most $dn$ iterations of $\mathcal{A}$, where $d$ is the minimum T-depth of $U$, so space and time complexity is $\poly\left(d,n,2^{5.6n}\right)$.

If we assume the weaker Conjecture \ref{conj:minDepth2} then we get a space and time complexity $\poly\left(d ,n^{\log n},2^{5.6n}\right)$.

\subsection{Implementations and results}

\begin{table}
 \centering
 \footnotesize
 \begin{tabular}{|l|c|c|c|c|c|c|c|}
 \hline
 Unitary & $\#$qubits & T-depth & T-count & Optimal? & Time & Max $\#$nodes & Prev T-depth\\
\hline 
Toffoli & 3 & 3 & 7 & Yes & 27m 41s & 358  & 3 \cite{2013_AMMR}  \\ 
\hline
Fredkin & 3 & 3 & 7 & Yes & 29m 49s & 386 & 4 \cite{2013_AMMR}\\
\hline
Peres & 3 & 3 & 7 & Yes & 27m 36s & 358 & 4\cite{2013_AMMR} \\
\hline
Quantum OR & 3 & 3 & 7 & Yes & 27m 35s & 358 & 4\cite{2013_AMMR}\\
\hline
Negated Toffoli & 3 & 3 & 7 & Yes & 27m 12s & 358 & 4\cite{2013_AMMR}  \\
\hline
 \end{tabular}
 \caption{Performance of our algorithm on some benchmark circuit unitaries. The T-depth returned by our algorithm (3rd column) is optimal for all unitaries. In most cases it is less than the T-depth of the circuits shown in \cite{2013_AMMR}. We have also tabulated the T-count (4th column) of the T-depth-optimal circuits, running time (5th column) as well as the maximum number of nodes or intermediate unitaries (6th column) that accumulate at any level while running our algorithm. The running time excludes the pre-processing time to generate $\mathbb{V}_3$.  }
 \label{tab:benchmark}
\end{table}

\begin{table}[!htbp]
\centering
\footnotesize
\begin{tabular}{|c|c|c|c|c|c|}
 \hline
 $\#$Qubits & Max. T-depth & Time (avg) & Time (std) & Max. $\#$nodes (avg) & Max. $\#$nodes (std)\\
 \hline\hline
 \multirow{11}{*}{2} & 2 & 0.015s & 0.006s & 2.90 & 3.45\\
 \cline{2-6}
 & 3 & 0.055s & 0.035s & 6.70 & 4.75  \\
 \cline{2-6}
 & 4 & 0.184s & 0.153s & 21.1 & 20.1 \\
 \cline{2-6}
 & 5 & 0.49s & 0.62s & 56.5 & 71.7 \\
 \cline{2-6}
 & 6 & 1.17s & 0.91s & 99.9 & 73.0 \\
 \cline{2-6}
 & 7 & 3.10s & 4.19s & 256.3 & 355.0 \\
 \cline{2-6}
 & 8 & 8.39s & 8.36s & 443.0 & 435.2 \\
 \cline{2-6}
 & 9 & 15.1s & 7.81s & 721.1 & 499.9 \\
 \cline{2-6}
 & 10 & 38.1s & 32.5s & 1727.7 & 1282.7 \\
 \cline{2-6}
 & 11 & 48.3s & 50.5s & 2049.7 & 1809.5 \\
 \cline{2-6}
 & 12 & 47.6s & 59.5s & 1853.9 & 2218.8 \\
 \cline{2-6}
 & 13 & 188.2s & 158.8s & 6705.8 & 5991.0 \\
 \cline{2-6}
 & 14 & 547.1s & 921.0s & 12293.9 & 16351.1 \\
 \cline{2-6}
 & 15 & 315.8s & 295.4s & 9515.6 & 7925.2 \\
 \cline{2-6}
 & 16 & 238.3s & 169.1s & 7025.4 & 4905.8 \\
 \cline{2-6}
 & 17 & 495.8s & 589.7s & 12118.6 & 13545.6 \\
 \cline{2-6}
 & 18 & 408.3s & 265.3s & 9466.6 & 4937.4 \\
 \cline{2-6}
 & 19 & 625.8s & 478.1s & 14390.3 & 9332.9 \\
 \cline{2-6}
 & 20 & 1008.2s & 656.6s & 15313.2 & 9205.1 \\
 \hline\hline
 \multirow{5}{*}{3} & 2 & 17.4s & 21.6s & 8.50 & 8.36 \\
 \cline{2-6}
 & 3 & 209.4s & 190.8s & 123.5 & 173.0 \\
 \cline{2-6}
 & 4 & 999.9s & 780.3s & 253.2 & 226.2 \\
 \cline{2-6}
 & 5 & 3926.6s & 3424.7s & 1203.5 & 1968.8 \\
 \cline{2-6}
 & 6 & 11349.5s & 11076.0s & 1024.1 & 643.1 \\
  \cline{2-6}
 & 7 & 28750.9s & 18652.0s & 4481.3 & 4165.9 \\
 \hline
\end{tabular}
\caption{Performance of MIN-T-DEPTH on random circuits. For each entry in the table, we generate 10 random circuits.}
\label{tab:random}
\end{table}

We implemented our algorithm MIN-T-DEPTH in standard C++17 on an Intel(R) Core(TM) i7-7700K CPU at 4.2GHz, with 8 cores and 16 GB RAM, running Debian Linux 9.13. We used OpenMP~\cite{OpenMP}
for parallelization  and the Eigen~3 matrix library~\cite{Eigen} for some of the matrix operations. Our algorithm returns a T-depth-optimal decomposition of an input unitary. We can generate a circuit for each $R(P)$ using Fact \ref{fact:cliffConj} in Section \ref{app:prelim:cliffPauli} and the trailing Clifford using the algorithm in \cite{2004_AG}. We remind the reader that the numerical results of this subsection, together with instructions on how to reproduce them, are available online at \url{https://github.com/vsoftco/t-depth}. We have implemented MIN-T-DEPTH and not the optimal nested MITM algorithm in Section \ref{provable:Tdepth} because the former has better complexity.

We have synthesized T-depth-optimal circuits for 3 qubit benchmark unitaries like Toffoli, Fredkin, Peres, Quantum OR, Negated Toffoli (Table \ref{tab:benchmark}). We found the min-T-depth of all these unitaries is 3, which is less than the T-depth of the circuits shown in \cite{2013_AMMR} (except Toffoli). The authors did not perform a T-depth-optimal synthesis of these 3 qubit circuits, since their algorithm required to generate a (pre-processed) set of more than 92,897,280 elements, which took more than 4 days (Table \ref{tab:Vn}). The running time as well as space requirement, being an exponential (in min-T-depth) of this set, it would have been intractable on a PC. The largest T-depth-optimal circuit implemented in \cite{2013_AMMR} had 2-qubits and had T-depth 2.
In our case the set generated during pre-processing is $\mathbb{V}_n$. In case of 3 qubits it has 2282 elements and takes about 2 seconds to be generated. The average searching time is 27.5 minutes. Thus our algorithm clearly outperforms the previously best T-depth-optimal synthesis algorithm in \cite{2013_AMMR}. 

We would like to mention here that for T-depth-optimal synthesis algorithms like \cite{2013_AMMR} or ours, the input is a unitary matrix and no other additional information is provided. The T-depth of some unitaries may be related. For example, the authors have been pointed out that T-depth of Fredkin, Peres can be obtained from T-depth of Toffoli because they are Clifford equivalent. There are some concerns here. We do not know of any efficient test for Clifford equivalence given arbitrary exactly implementable unitaries. Second, we are unaware of any set of benchmark unitaries from which we can derive the T-depth of any exactly implementable unitary. In fact, these extra information can serve as litmus tests for the correctness of the output of any algorithm. 

We have synthesized T-depth-optimal circuits for 2 and 3-qubit permutation unitaries. We found that all 2-qubit permutations are Cliffords. It took us, on an average, 0.726 seconds to synthesize 2-qubit permutations. We considered about 100 random 3-qubit permutation unitaries and (due to time constraint) we synthesized completely (up to Clifford) the unitaries with T-depth at most 5. The permutations with T-depth at most 3 took on average 15 mins. The permutations with T-depth at most 5 took on average 4.5 hours. 

We have also tested our algorithm on random 2 and 3 qubit circuits (Table~\ref{tab:random}). The input 2 and 3 qubit circuits had T-depth 2-10 and 2-7, respectively. Each line in Table~\ref{tab:random} is computed from 10 random circuits. By Max.$\#$ nodes we mean the maximum number of unitaries selected at any level. ``avg'' means we average this statistic over all unitaries considered. ``std'' means we find the standard deviation of this statistic.
We found out that the circuits output by our algorithm had T-depth at most the input T-depth. Now the min-T-depth can be at most the input T-depth. We could not verify the optimality of our results, since we do not know of any T-depth-optimal synthesis algorithm that can implement such large circuits. However, this is good indication that our algorithm MIN-T-DEPTH actually obtains the min-T-depth for most unitaries.

\section{Conclusion}
\label{sec:conclude}

We study the complexity of synthesizing T-depth optimal circuits of exactly implementable unitaries. First, we define a subset, $\mathbb{V}_n$, of T-depth-1 unitaries that have certain properties, with which we can generate the T-depth-optimal decomposition of any exactly synthesizable unitary (up to a Clifford). We prove $|\mathbb{V}_n|\in O(n\cdot 2^{5.6n})$.

We adopt the nested meet-in-the-middle (MITM) technique used by Mosca and Mukhopadhyay~\cite{2020_MM} to synthesize T-count optimal circuits.
First we show a space-time trade-off for the application of MITM techniques to the depth-optimal synthesis problem.
To synthesize T-depth optimal circuits we further use the channel representation of unitaries and $\mathbb{V}_n$ in the nested MITM framework. 
We achieve a space and time complexity $\exp\left(4^{n^2}, d\right)$. The T-depth-optimal synthesis algorithm in \cite{2013_AMMR} had a space and time complexity $\exp\left(3^n|\cliff_n|,d\right)$, where $\cliff_n$ is the set of n-qubit Clifford operators (up to global phase). Since $|\cliff_n|\in O\left(2^{kn^2}\right)$ ($k>2.5$), we achieve a significant speed-up.

To increase the efficiency further we develop another algorithm that returns a T-depth-optimal decomposition with complexity $\poly(n,2^{5.6n},d)$ if some conjectures are true. Assuming a weaker conjecture this complexity is $\poly(n^{\log n},2^{5.6n},d)$. Using this algorithm, we could synthesize T-depth-optimal circuits of larger unitaries than in \cite{2013_AMMR} in reasonable time. 

\section*{Acknowledgement}
The authors wish to thank NTT Research for their financial and technical support. This work was supported in part by Canada's NSERC.  IQC and the Perimeter Institute (PI) are supported in part by the Government of Canada and Province of Ontario (PI). We thank the anonymous reviewers for their comments, that not only helped us improve the write-up significantly, but also led to a tighter bound in Fact 3.1. 

\section*{Author contributions}

The ideas were given by P.Mukhopadhyay. The software implementations were done by V. Gheorghiu. All the authors contributed to the preparation of the manuscript. 

\section*{Data availability}

Numerical results together with instructions on how to reproduce them, are available online at 
\url{https://github.com/vsoftco/t-depth}.

\section*{Code availability}

The code is available from the corresponding author on request.

\section*{Competing interests}

Michele Mosca is co-founder of softwareQ Inc. and has filed a provisional patent application for this work. P.Mukhopadhyay is a co-inventor in this patent. 


\appendix
\section{Some additional preliminaries}
\label{app:prelim}

\subsection{Cliffords and Paulis}
\label{app:prelim:cliffPauli}

The \emph{single qubit Pauli matrices} are as follows:
\begin{eqnarray}
 \X=\begin{bmatrix}
     0 & 1 \\
    1 & 0
    \end{bmatrix} \qquad  
 \Y=\begin{bmatrix}
     0 & -i \\
     i & 0
    \end{bmatrix} \qquad 
 \Z=\begin{bmatrix}
     1 & 0 \\
     0 & -1
    \end{bmatrix}
\label{eqn:Pauli1}
\end{eqnarray}

The \emph{$n$-qubit Pauli operators} are :
\begin{eqnarray}
 \pauli_n=\{Q_1\otimes Q_2\otimes\ldots\otimes Q_n:Q_i\in\{\id,\X,\Y,\Z\} \}.
 \label{eqn:paulin}
\end{eqnarray}

The \emph{single-qubit Clifford group} $\cliff_1$ is generated by the Hadamard and phase gate.
\begin{eqnarray}
 \cliff_1=\braket{\had,\phase} 
 \label{eqn:cliff1}
\end{eqnarray}
where
\begin{eqnarray}
 \had=\frac{1}{\sqrt{2}}\begin{bmatrix}
       1 & 1 \\
       1 & -1
      \end{bmatrix}\qquad 
 \phase=\begin{bmatrix}
       1 & 0 \\
       0 & i
      \end{bmatrix}.
\end{eqnarray}
When $n>1$ the \emph{$n$-qubit Clifford group} $\cliff_n$ is generated by these two gates (acting on any of the $n$ qubits) along with the two-qubit $\CNOT=\ket{0}\bra{0}\otimes\id+\ket{1}\bra{1}\otimes\X$ gate (acting on any pair of qubits). 

Cliffords map Paulis to Paulis up to a possible phase of $-1$ i.e. for any $P\in\pauli_n$ and any $C\in\cliff_n$ we have
$$
    CPC^{\dagger}=(-1)^bP'
$$
for some $b\in\{0,1\}$ and $P'\in\pauli_n$. In fact, given two Paulis (neither equal to the identity), it is always possible to efficiently find a Clifford which maps one to the other.
\begin{fact}[\cite{2014_GKMR}]
 For any $P,P'\in\pauli_n\setminus\{\id\} $ there exists a Clifford $C\in\cliff_n$ such that $CPC^{\dagger}=P'$. A circuit for $C$ over the gate set $\{\had,\phase,\CNOT\}$ can be computed efficiently (as a function of $n$).
 \label{fact:cliffConj}
\end{fact}

\subsection{The group $\mathcal{J}_n$ generated by Clifford and $\T$ gates}

The group $\mathcal{J}_n$ is generated by the $n$-qubit Clifford group along with the $\T$ gate, where
\begin{eqnarray}
 \T=\begin{bmatrix}
     1 & 0 \\
     0 & e^{i\frac{\pi}{4}}
    \end{bmatrix}
\end{eqnarray}

Thus for a single qubit
\begin{eqnarray}
 \mathcal{J}_1 = \braket{\had,\T}    \nonumber
\end{eqnarray}
and for $n>1$ qubits
\begin{eqnarray}
 \mathcal{J}_n=\braket{\had_{(i)},\T_{(i)},\CNOT_{(i,j)}:i,j\in [n]}.
 \nonumber
\end{eqnarray}
It can be easily verified that $\mathcal{J}_n$ is a group, since the $\had$ and $\CNOT$ gates are their own inverses and $\T^{-1}=\T^7$. Here we note $\phase=\T^2$.

We denote the group of unitaries exactly synthesized over the Clifford+T gate set by $\clifft_n$. Some elements of this group cannot be exactly synthesized over this gate set without ancilla qubits. 

The following characterization of this group was proved by Giles and Selinger \cite{2013_GS}.

\begin{theorem} (Theorem 1 and Corollary 2 from \cite{2013_GS})
Let $U(N)$ (where $N=2^n$) is the group of $n$-qubit unitaries. Then the following are equivalent.
\begin{enumerate}
 \item $U$ can be exactly represented by a quantum circuit over the Clifford+T gate set, possibly using some finite number of ancillas that are initialized and finalized in state $\ket{0}$.
 
 If ancilla is required, then a single ancilla is sufficient.
 
 \item The entries of $U$ belong to the ring $\intg\left[i,\frac{1}{\sqrt{2}}\right]$.
where 
 \begin{eqnarray}
 \intg\left[i,\frac{1}{\sqrt{2}}\right]&=&\Big\{(a+bi+c\sqrt{2}+di\sqrt{2})/\sqrt{2}^k  \nonumber \\
 &&:a,b,c,d\in\intg,\quad k\in\nat \Big\}.  \nonumber
\end{eqnarray}
 
 \item No ancilla is required if $\det(U)=e^{i\frac{\pi}{8}Nr} $ for some $r\in[8]$. For $n\geq 4$ the condition on the determinant is simply $\det(U)=1$.
\end{enumerate}
 \label{thm:2013_GS}
\end{theorem}

\subsection{Channel representation}
\label{app:chan}

\begin{lemma}[\cite{2020_MM}]
Let $U$ and $V$ are $N_1\times N_1$ and $N_2\times N_2$ unitaries, where $N_1=2^{2n}$ and $N_2=2^{2m}$. Then
$
    \chan{(V\otimes U)}=\chan{V}\otimes\chan{U}.
$
 \label{lem:chanAncilla2}
\end{lemma}

Let $U\in\clifft_n$ and $\ket{\phi}$, $\ket{\psi}$ are the ancilla and input state respectively. To be specific, $U'$ is the unitary that acts on the joint state space of ancilla and input qubits. For tensor product inputs, we have
$$
    U'(\ket{\phi}\otimes\ket{\psi})=V\ket{\phi}\otimes U\ket{\psi}=(V\otimes U)(\ket{\phi}\otimes\ket{\psi})
$$
Since the product states span the entire state space, we must have $U'=V\otimes U$. So from Lemma \ref{lem:chanAncilla2} we can calculate $\chan{U'}$ from $\chan{U}$ and $\chan{V}$.
$$
    \chan{U'}=\chan{V}\otimes\chan{U}
$$
Note in many cases the ancilla remains unchanged at the end of operations i.e. $V=\id$.
From here on, with a slight abuse of notation when we write $U\in\clifft_n$ we assume it is the unitary that acts on the joint state space of input and ancilla qubits. 

\begin{lemma}
If $U=e^{i\phi}\left(\prod_{j=t}^1\widetilde{R}(P_j)\right)C_0$ where $\widetilde{R}\in\{R,R^{\dagger}\}$, $P_j\in\pauli_n\setminus\{\id\}$ and $C_0\in\cliff_n$, then
$$
    \id_m\otimes U=e^{i\phi}\left(\prod_{j=t}^1\widetilde{R}(\id_m\otimes P_j)\right)(\id_m\otimes C_0)
$$
Also
$$
    \widetilde{R}(\id_m\otimes P)=\id_m\otimes\widetilde{R}(P)
$$
\label{lem:decomposeAncilla} 
\end{lemma}

\begin{proof}
 We know $R(P)=\frac{1}{2}\left(1+e^{i\frac{\pi}{4}}\right)\id_n+\frac{1}{2}\left(1-e^{i\frac{\pi}{4}}\right)P = \alpha\id_n+\beta P$ (Let). Then
 \begin{eqnarray}
  R(\id_m\otimes P)&=&\alpha(\id_m\otimes\id_n)+\beta(\id_m\otimes P) = (\id_m\otimes\alpha\id_n)+(\id_m\otimes\beta P) \nonumber   \\
  &=& \id_m\otimes(\alpha\id_n+\beta P) = \id_m\otimes R(P) \nonumber
 \end{eqnarray}
Similarly we can show that $R^{\dagger}(\id_m\otimes P)=\id_m\otimes R^{\dagger}(P)$. Now
\begin{eqnarray}
 \id_m\otimes U &=& e^{i\phi}\left(\id_m\otimes\left(\prod_{j=t}^1\widetilde{R}(P_j)\right)C_0\right) \nonumber \\
 &=& e^{i\phi}\left(\prod_{j=t}^1\left(\id_m\otimes\widetilde{R}(P_j)\right)\right)\left(\id_m\otimes C_0\right)  \nonumber \\
 &=& e^{i\phi}\left(\prod_{j=t}^1 \widetilde{R}(\id_m\otimes P_j)\right)(\id_m\otimes C_0) \nonumber
\end{eqnarray}
\end{proof}

Thus from Lemma \ref{lem:decomposeAncilla} we get one decomposition of $U'=\id_m\otimes U$ and hence $\chan{U'}$.
\begin{eqnarray}
\chan{U'}= \chan{\id_m\otimes U}=
\left(\prod_{j=t}^1\chan{\widetilde{R}(\id_m\otimes P_j)}\right)(\chan{\id_m}\otimes\chan{C_0})
\label{app:eqn:decomposeAncilla}
\end{eqnarray}

\subsection{Properties of $\chan{R(P)}$ and $\chan{R^{\dagger}(P)}$ for $n$-qubit non-identity Paulis}
\label{app:RP}

The following facts about the structure of $\chan{R(P)}$, where $P\in\pauli_n\setminus\{\id\}$, have been proved in \cite{2020_MM}. 
\begin{lemma}
 \begin{enumerate}
  \item The diagonal entries of $\chan{R(P)}$ are $1$ or $\frac{1}{\sqrt{2}}$.
  
  \item If a diagonal entry is $1$ then all other entries in the corresponding row and column is $0$.
  
  \item If a diagonal entry is $\frac{1}{\sqrt{2}}$ then one other entry in the corresponding row is $\pm\frac{1}{\sqrt{2}}$ and one other entry in the corresponding column is $\mp\frac{1}{\sqrt{2}}$.
  
  \item Exactly $2^{2n-1}$ diagonal elements can be $\frac{1}{\sqrt{2}}$.
 \end{enumerate}

\end{lemma}

$\chan{R^{\dagger}(P)}=\chan{R(P)}^{\dagger}=\chan{R(P)}^{-1}$ and the above properties also hold for $\chan{R^{\dagger}(P)}$. $\chan{R(P)}$ can be represented compactly by an array of size $2^{2n-2}$ and from this $\chan{R^{\dagger}(P)}$ can be computed very efficiently \cite{2020_MM}.

An $O\left(N^4\right)$ time algorithm for multiplying two $N^2\times N^2$ unitaries $\chan{\widetilde{R}(P)}$ and $W$ (where $N=2^n$) has been given in \cite{2020_MM}. To the best of our knowledge, currently the fastest algorithm \cite{2014_LG, 2021_AW} for matrix multiplication incurs a time complexity of $O(N^{4.7457278})$ while multiplying two $N^2\times N^2$ matrices.


\section{ Proof of Lemma \ref{lem:provNest} }
\label{sec:appendix}

Now we prove Lemma \ref{lem:provNest} in Section \ref{sec:provable}.

\begin{lemma}
Let $S_i\subset U(2^n)$ be the set of all unitaries implementable in depth $i$ over the gate set $\mathcal{G}$. Given a unitary $U$, there exists a circuit over $\mathcal{G}$ of depth $(d_1+d_2)$ implementing $U$ if and only if $S_{d_1}^{\dagger}U\bigcap S_{d_2}\neq\emptyset$.
 \label{app:lem:provNest}
\end{lemma}

\begin{proof}
 We note that $U\in S_i^{\dagger}=\{U^{\dagger}|U\in S_i \}$ if and only if $U$ can be implemented in depth $i$ over $\mathcal{G}$. (Though this was proved in Lemma 1 of \cite{2013_AMMR} we include it briefly here for completion.) Let $U=U_1U_2\ldots U_i$ where $U_1,U_2,\ldots,U_i\in\mathcal{V}_{n,\mathcal{G}}$ and so $U^{\dagger}=U_i^{\dagger}\ldots U_2^{\dagger}U_1^{\dagger}$. As $\mathcal{G}$ is closed under inversion so $U_1^{\dagger},U_2^{\dagger},\ldots,U_i^{\dagger}\in\mathcal{V}_{n,\mathcal{G}}$, and thus a circuit of depth $i$ over $\mathcal{G}$ implements $U^{\dagger}$. Since $\left(S_i^{\dagger} \right)^{\dagger}=S_i$ the reverse direction follows.
 
 Suppose $U$ is implementable by a circuit $C$ of depth $d_1+d_2$. We divide $C$ into two circuits of depth $d_1$ and $d_2$, implementing unitaries $W_1\in S_{d_1}$ and $W_2\in S_{d_2}$ respectively, where $W_1W_2=U$. So $W_2=W_1^{\dagger}U \in S_{d_1}^{\dagger}U$ and hence $W_2\in S_{d_1}^{\dagger}U \bigcap S_{d_2}$.
 
 In the other direction let $S_{d_1}^{\dagger}U\bigcap S_{d_2}\neq\emptyset$. So there exists some $W_2\in S_{d_1}^{\dagger}U\bigcap S_{d_2}$. Since $W_2\in S_{d_1}^{\dagger}U$ so $W_2=W_1^{\dagger}U$ for some $W_1\in S_{d_1}^{\dagger}$. Now $W_2\in S_{d_2}$ and $W_1W_2=U$. Thus $U$ is implementable by some circuit of depth $d_1+d_2$.
\end{proof}

\section{Pseudocode of algorithms}
\label{sec:app:code}

In this section we give the pseudocode for the nested MITM algorithm ( Section \ref{sec:provable}).

\begin{algorithm}
\scriptsize
 \caption{Nested MITM}
 \label{app:alg:nestMITM}
 \KwIn{(i) A unitary $U$, (ii) a gate set $\mathcal{G}$, (iii) depth $d$, (iv) $c\geq 2$}
 \KwOut{A circuit (if it exists) for $U$ such that depth is at most $d$. }
 
 Generate the set $\mathcal{V}_{n,\mathcal{G}}$ of $n$-qubit unitaries with depth 1. \; \label{nestMITM:generate}
 $S_0 \leftarrow \{\id\}$;  $\quad i\leftarrow 1$     \;  \label{nestMITM:start}
 \While{$i\leq \left\lceil\frac{d}{c}\right\rceil$ \label{nestMITM:whileStart}}
 {
    $S_i\leftarrow \mathcal{V}_{n,\mathcal{G}}S_{i-1} $ \; \label{nestMITM:depthInc}
    $k=c-1$ \;\label{nestMITM:forStart}
    
        \For{$W=W_1W_2\ldots W_k$ where $W_i\in S_i$ or $W_i\in S_{i-1}$ \label{nestMITM:for}}
        {
            \If{$\exists W'\in S_i$ such that $W^{\dagger}U=W'$ \label{nestMITM:check1}}
            {
                  \Return $W_1,W_2,\ldots,W_k,W'$  \label{nestMITM:return1}\;  
                   break \; 
            }
            \ElseIf{$\exists W'\in S_{i-1}$ such that $W^{\dagger}U=W'$ \label{nestMITM:check2}}
            {
                  \Return $W_1,W_2,\ldots,W_k,W'$  \label{nestMITM:return2}\;  
                   break \; 
            }
        }\label{nestMITM:forEnd}
   $i\leftarrow i+1$ \;
 }  \label{nestMITM:whileEnd}
 \If{no decomposition found}
 {
    \Return "$U$ has depth more than $d$."  \;
 }
 
\end{algorithm}

Now we give the pseudocode for the MIN T-DEPTH algorithm and its sub-procedure $\mathcal{A}$ (Section \ref{sec:minDepthHeuristic}).

\begin{algorithm}
\scriptsize
 \caption{$\mathcal{A}$}
 \label{app:alg:Aheuristic}
 
 \KwIn{(i) $\chan{U}$ where $U\in\clifft_n$ and (ii) integer $d' > 0$}
 \KwOut{Set $\mathcal{D}$ of decompositions $\chan{U}=\left(\prod_{i=k}^1\chan{R(P)}\right)\chan{C_0}$ if minimum T-depth at most $d'$}
 
 Compute sets $\mathbb{V}_{n,j}^{-1}$ for $j=1,2,\ldots n$. \tcp*[f]{Can do pre-processing} \;
 $\upath_{\chan{U}}\leftarrow []$; $\mathcal{D}\leftarrow\emptyset$ \; 
 $\widetilde{U}=[\chan{U},\upath_{\chan{U}},\sde_{\chan{U}},\ham_{\chan{U}}]$   \tcp*[f]{Stores root node} \;
 $\mathcal{U}_0=\{\widetilde{U}\}$ \tcp*[f]{Subscript indicates path T-count} \;
 \For{$i=1,2,\ldots,d'$}
 {
    $SH\leftarrow\emptyset$ \tcp*[f]{Stores tuples (sde, Hamming weight change, path T-count, count)} \;
    $P\leftarrow\emptyset$ \tcp*[f]{Stores the path T-counts obtained}  \;
    \For{$k=i-1,\ldots,n(i-1)$ \tcp*[h]{Parent hypernodes//}}
    {
        \For{$\widetilde{U}\in\mathcal{U}_k$ \tcp*[h]{Each node within a hypernode//}}
        {
            \For{$j=1,2,\ldots,n$}
            {
                \For{each $\chan{V}^{-1}\in \mathbb{V}_{n,j}^{-1}$}
                {
                    $\chan{W}\leftarrow\chan{V}^{-1}\chan{U}$   \;
                    $\Delta\ham_{\chan{W}} = $ inc, dec or same if $\ham_{\chan{W}} > \ham_{\chan{U}}$, $\ham_{\chan{W}} < \ham_{\chan{U}}$ or $\ham_{\chan{W}} = \ham_{\chan{U}}$ respectively \;
                    $P.append(PathT_{\chan{W}})$ if $PathT_{\chan{W}}\notin P$  \tcp*[f]{$PathT_{\chan{W}}$} is the Path T-count\;
                    \eIf{$(\sde_{\chan{W}},\Delta\ham_{\chan{W}},PathT_{\chan{W}},count) \in SH$}
                    {
                        $(\sde_{\chan{W}},\Delta\ham_{\chan{W}},PathT_{\chan{W}},count)\leftarrow (\sde_{\chan{W}},\Delta\ham_{\chan{W}},PathT_{\chan{W}},count+1) $  \;
                    }
                    {  
                        $SH.append((\sde_{\chan{W}},\Delta\ham_{\chan{W}},PathT_{\chan{W}},1))$ \;
                    }
                }
            }
        }
    }
    \For{$p\in P$}
    {
        $(s,\Delta h, p,count)\leftarrow$ Element of $SH$ such that $count$ is minimum, 
        $s\leq d'-i-1$ and Path T-count is $p$ \; \label{Aheuristic:takeMin}
        $G\leftarrow$ Product unitaries such that $\sde_{\chan{W}}=s$, $\Delta\ham_{\chan{W}}=\Delta h$ and 
        Path T-count is $p$  \tcp*[f]{Can be obtained by re-computing those unitaries} \;
    }
    \For{$\chan{U'}\in G$ }
    {
        \eIf{$sde_{\chan{U'}}==0$}
        {
            $\chan{C_0}=\chan{U'}$ \;
            \textbf{Store} $\upath_{\chan{U'}}, \chan{C_0}$ as a decomposition of $\chan{U}$ and $i$ as its depth in $\mathcal{D}$    \;
        }
        {
            $\widetilde{U'}=[\chan{U'},\upath_{\chan{U'}},\sde_{\chan{U'}},\ham_{\chan{U'}}]$  \tcp*[f]{Build children node} \;
            $\mathcal{U}_k$.append$(\widetilde{U'})$, where $k$ is the path T-count of $\chan{U'}$  \tcp*[f]{Store in hypernodes}   \;
        }
    }
    Delete all parent nodes;
 }
 \Return $\mathcal{D}$    \;
\end{algorithm}

\begin{algorithm}
\scriptsize
\caption{MIN T-DEPTH}
\label{app:alg:minTdepth}
 
 \KwIn{$\chan{U}$ where $U\in\clifft_n$}
 \KwOut{$d$, the minimum T-depth of $U$ and its decomposition $\chan{U}=\left(\prod_{i=k}^1\chan{R(P)} \right)\chan{C_0}$}
 
 $d'=\left\lceil\frac{\sde_{\chan{U}}}{n}\right\rceil$ or $\left\lceil\frac{\tcount(U)}{n}\right\rceil$ if $\tcount(U)$ is known\;
 \While{$1$}
 {
    $\mathcal{A}$($\chan{U},d'$)   \;
    \eIf{$\mathcal{D}==\emptyset$}
    {
        $d'\leftarrow d'+1$ \;
    }
    {
        In each decomposition check if consecutive unitaries can be combined to form a T-depth 1 product \;
        \Return a decomposition with the minimum T-depth    \;
        break   \;
    }
 }
\end{algorithm}

\end{document}